\documentclass[11pt]{article}
\pdfoutput=1
\usepackage{pifont,cancel,comment}
\usepackage[utf8]{inputenc}
\usepackage{here}
\usepackage{subfigure}
\usepackage{wrapfig}
\usepackage{subfiles}
\usepackage{graphicx,xcolor}
\usepackage{amsmath,amssymb,amsbsy,amstext,amsthm,simplewick,amsfonts}
\usepackage{feynmp-auto}
\usepackage{slashed}
\usepackage{lmodern}
\usepackage{bm}
\usepackage{caption}
\usepackage{subcaption}
\usepackage{physics}
\usepackage{braket}
\usepackage[version=4]{mhchem}
\usepackage{wasysym}
\usepackage{upgreek}
\usepackage{textgreek}

\usepackage{cite}

\usepackage{hyperref}
\hypersetup{colorlinks=true,linkcolor=bluecyan,citecolor=red3,urlcolor=green3,pdfencoding=auto,linktocpage}

\usepackage{tikz}
\usetikzlibrary{graphs}

\usepackage[framemethod=default]{mdframed}
\newmdenv[backgroundcolor=gray!15,%
skipabove=5pt,%
skipbelow=5pt,%
leftmargin=2pt,%
rightmargin=2pt,%
innertopmargin=-6pt,%
innerbottommargin=5pt,%
innerleftmargin=5pt,%
innerrightmargin=5pt,%
splittopskip=0pt,%
splitbottomskip=0pt,%
linewidth=0pt,%
nobreak=true]%
{keyeqn}

\newmdenv[backgroundcolor=gray!15,%
skipabove=5pt,%
skipbelow=5pt,%
leftmargin=2pt,%
rightmargin=2pt,%
innertopmargin=-2pt,%
innerbottommargin=5pt,%
innerleftmargin=5pt,%
innerrightmargin=5pt,%
splittopskip=0pt,%
splitbottomskip=0pt,%
linewidth=0pt,%
nobreak=true]%
{keythrm}

\graphicspath{{Fig/}}

\definecolor{lightgreen}{cmyk}{0.2, 0, 0.2, 0.2}
\definecolor{lightgray}{cmyk}{0.1,0.2,0,0.1}
\definecolor{lightgray2}{cmyk}{0.1,0.1,0,0.1}
\definecolor{greyish2}{rgb}{.96,.96,.96}
\definecolor{bluecyan}{RGB}{0, 100, 200}
\definecolor{blue3}{RGB}{31,119,180}
\definecolor{red3}{RGB}{214,39,40}
\definecolor{orange3}{RGB}{255,127,14}
\definecolor{green3}{RGB}{44,160,44}
\definecolor{red2}{RGB}{255,0,0}
\definecolor{green2}{RGB}{0,170,0}
\definecolor{blue2}{RGB}{0,128,255}
\definecolor{magenta2}{RGB}{191,64,191}
\definecolor{purple2}{RGB}{112,48,160}
\definecolor{orange2}{RGB}{255,192,0}
\definecolor{blue2}{RGB}{117,223,230}

\newcommand{\email}[1]{\footnote{\href{mailto:#1}{#1}}}

\numberwithin{equation}{section}

\allowdisplaybreaks[1]
\renewcommand{\eqref}[1]{(\ref{#1})}

\renewcommand{\i}{\mathrm{i}}

\makeatletter

\@addtoreset{equation}{section}

\newcommand{\lambdabar}{{\mathchoice
  {\smash@bar\textfont\displaystyle{0.25}{1.2}\lambda}
  {\smash@bar\textfont\textstyle{0.25}{1.2}\lambda}
  {\smash@bar\scriptfont\scriptstyle{0.25}{1.2}\lambda}
  {\smash@bar\scriptscriptfont\scriptscriptstyle{0.25}{1.2}\lambda}
}}
\newcommand{\smash@bar}[4]{
  \smash{\rlap{\raisebox{-#3\fontdimen5#10}{$\m@th#2\mkern#4mu\mathchar'26$}}}%
}
\makeatother

\usepackage[whole]{bxcjkjatype}
\usepackage{mathtools}
\usepackage{array}
\usepackage{tikz}
\usepackage{bm}
\usepackage[T1]{fontenc}
\usepackage[normalem]{ulem}
\usetikzlibrary{arrows.meta}
\begin{document}
\begin{titlepage}
\setcounter{page}{1}
\baselineskip=15.5pt
\thispagestyle{empty}
\begin{center}
{\fontsize{18}{22}\selectfont\bfseries
Warm Inflation with Environmental Backreaction from Schwinger--Keldysh Effective Field Theory\par}
\end{center}
\vskip 15pt
\begin{center}
{\fontsize{12}{18}\selectfont
Kanji Nishii$^{a,b,}$\email{kanji.nishii@keio.jp},
Toshifumi Noumi$^{c,d,}$\email{tnoumi@g.ecc.u-tokyo.ac.jp},
and Fumiya Sano$^{c,}$\email{sanof.cosmo@gmail.com}\par}
\end{center}
\begin{center}
\vskip 8pt
$^a$ \textit{Graduate School of Science and Technology, Keio University,}\\
\textit{Yokohama, Kanagawa 223-8522, Japan}\\[5pt]
$^b$ \textit{Quantum Computing Center, Keio University,}\\
\textit{3-14-1 Hiyoshi, Kohoku-ku, Yokohama, Kanagawa 223-8522, Japan}\\[5pt]
$^c$ \textit{Graduate School of Arts and Sciences, The University of Tokyo,}\\
\textit{Komaba, Meguro-ku, Tokyo 153-8902, Japan}\\[5pt]
$^d$ \textit{RIKEN Center for Interdisciplinary Theoretical and Mathematical Sciences (iTHEMS),}\\
\textit{RIKEN, Wako 351-0198, Japan}
\end{center}
\vspace{0.4cm}
\begin{center}\textbf{Abstract}\end{center}

We construct a Schwinger--Keldysh effective field theory for warm inflation, modeling the thermal environment based on the effective field theory of fluctuating hydrodynamics. Keeping the fluid dynamical allows us to incorporate local fluctuations of temperature, chemical potentials, and fluid velocity, as well as to describe energy transfer and dissipation in a generally covariant manner. The framework also includes charge-changing reactions, with applications to sphaleron-driven scenarios in mind. For illustration, we study a setup motivated by Standard Model warm inflation and quantify the effects of slow environmental modes on primordial perturbations. We find that chemical-potential fluctuations induce a modest shift in the scalar tilt, while sound modes have a more pronounced effect on both the tilt and the tensor-to-scalar ratio.

\end{titlepage}
\setcounter{page}{2}
\tableofcontents
\newpage
\setcounter{footnote}{0}

\section{Introduction}
\label{sec:introduction}
Cosmic inflation~\cite{Starobinsky:1980te,Guth:1980zm,Sato:1981qmu,Linde:1981mu,Albrecht:1982wi} provides a framework for explaining the homogeneity and isotropy of the Universe and the origin of primordial fluctuations, consistent with cosmological observations~\cite{Planck:2018jri,AtacamaCosmologyTelescope:2025nti}.
In its simplest realization, accelerated expansion is driven by the potential energy of a slowly rolling scalar field, the inflaton.
The microphysical origin of inflation and its connection to particle physics, however, remain open questions.
Moreover, inflation need not occur in isolation.
In warm inflation, energy transfer from the inflaton sustains a thermal bath during accelerated expansion and gives rise to additional friction in the inflaton dynamics~\cite{Berera:1995ie,Berera:2008ar}.

Minimal warm inflation~\cite{Berghaus:2019whh} realizes this scenario through an axion-like coupling of the inflaton $\phi$ to a non-Abelian gauge field, while avoiding large thermal corrections to the inflaton potential that can otherwise obstruct slowroll~\cite{Yokoyama:1998ju}:\footnote{
See also~\cite{Bastero-Gil:2016qru} for 
an earlier realization that suppresses the leading thermal corrections while maintaining dissipation.}
\begin{align}
 \mathcal L_{\mathrm{int}}
 &=\frac{\alpha_{\mathrm g}}{8\pi}\frac{\phi}{f}
 F^A_{\mu\nu}\widetilde F^{A\mu\nu},
 \label{eq:microscopic-axion-coupling}
\end{align}
where $F^A_{\mu\nu}$ is the gauge-field strength, $\widetilde{F}^{A\mu\nu}$ its dual, $A$ an adjoint index, $\alpha_{\mathrm g}=g_{\mathrm g}^2/(4\pi)$ the gauge fine-structure constant, and $f$ the inflaton decay constant.
The shift symmetry protects the inflaton potential from perturbative corrections, while sphaleron processes in the hot gauge plasma provide a source of dissipation.
These processes induce transitions between configurations with different Chern--Simons number.
A rolling inflaton biases the transitions, transferring energy to the plasma~\cite{McLerran:1990de}.
The original model considered a pure Yang--Mills plasma, and the mechanism has since been extended to Standard Model (SM) gauge interactions, particularly to quantum chromodynamics (QCD) plasma~\cite{Berghaus:2025dqi}.

The thermal environment affects not only the background evolution but also primordial perturbations.
Temperature and fluid-velocity fluctuations support sound waves, whose interaction with the inflaton can substantially modify the scalar power spectrum~\cite{Hall:2003zp,Graham:2009bf}.
Stochastic descriptions of the inflaton and radiation fluid have been developed~\cite{Bastero-Gil:2014jsa} and applied to sphaleron-driven warm inflation~\cite{Mirbabayi:2022cbt}, while recent studies follow the gauge-invariant evolution of inflaton and plasma perturbations through reheating~\cite{Laine:2024wyv,Laine:2025rll}.
In sphaleron-driven models with light fermions, topological transitions also generate axial charges.
The resulting chemical potentials oppose the bias induced by the rolling inflaton and modify the dissipation~\cite{McLerran:1990de,Drewes:2023khq}.
Gauge-helicity feedback was emphasized in Ref.~\cite{Broadberry:2025ggb}, and recent studies track the evolution of axial-charge and gauge-helicity perturbations alongside those of the inflaton and radiation~\cite{Dejrah:2026ltr}.
Moreover, the fluid response can retain memory on Hubble time scales\cite{Turiaci:2013dka}, so its effects cannot always be captured by an instantaneous friction coefficient.
These observations motivate a description that retains the slow environmental modes and their fluctuations as dynamical degrees of freedom.

In this paper, we extend the framework of Ref.~\cite{Lau:2024mqm} to construct a general Schwinger--Keldysh effective field theory (SK EFT) for warm inflation. We model the thermal environment by the EFT of fluctuating hydrodynamics \cite{Haehl:2015foa,Haehl:2015uoc,Crossley:2015evo,Glorioso:2017fpd,Jensen:2017kzi} (see also Ref.~\cite{Liu:2018kfw} for a review), retaining local fluctuations of temperature, chemical potentials, and fluid velocity.
We also include charge-changing reactions~\cite{Landry:2020tbh,Hongo:2024brb} to describe charge production and its backreaction on the inflaton and the thermal environment, with applications to sphaleron-driven scenarios in mind.
For illustration, we apply this framework to a setup motivated by Standard Model warm inflation and compute the primordial power spectra.
In particular, we find that chemical-potential fluctuations induce a modest shift in the scalar tilt, while sound modes have a more pronounced effect on both the tilt and the tensor-to-scalar ratio.
We also examine the conditions under which the environmental modes can be integrated out to obtain a local effective description of the adiabatic perturbation, keeping previous discussions on local EFT of dissipation and noise during inflation~\cite{LopezNacir:2011kk,Burgess:2014eoa,Creminelli:2023aly,Salcedo:2024smn} and discussions on memory effects in warm inflation~\cite{Gangopadhyay:2026vdu} in mind.

This paper is organized as follows.
In Sec.~\ref{sec:parent}, we construct the SK EFT and derive the constraints imposed by dynamical Kubo--Martin--Schwinger (KMS) symmetry~\cite{Glorioso:2017fpd}.
In 
Sec.~\ref{sec:cosmology}, we derive the inflationary background equations and the quadratic action for scalar perturbations.
In Sec.~\ref{sec:spectrum}, we compute the primordial power spectra and examine the conditions for a local effective description of the adiabatic perturbation.
The final section summarizes our conclusions.
Throughout the paper, we use the metric signature $(-,+,+,+)$ and natural units $\hbar=c=k_B=1$, with $M_{\mathrm{Pl}}$ denoting the reduced Planck mass.

\section{Schwinger--Keldysh EFT for warm inflation }
\label{sec:parent}

In this section, we construct a gravitational SK EFT for warm inflation with a fluid environment, extending the framework of Ref.~\cite{Lau:2024mqm}. See also Ref.~\cite{Turiaci:2013dka} for EFT using a fluid in warm inflation.
Retaining the fluid environment as a dynamical sector allows us to use local dissipation and noise kernels in SK EFT, as well as accounts for gauge invariance of the energy and momentum transfer from the inflaton.\footnote
{
Integrating out environmental degrees of freedom generally produces a nonlocal influence functional, in contrast to local EFT of dissipation and noise on inflaton in Refs.~\cite{LopezNacir:2011kk,Burgess:2014eoa,Creminelli:2023aly,Salcedo:2024smn}.
Moreover, constraints from long range interactions such as gauge forces can obstruct a naive local description of dissipation~\cite{Salcedo:2024nex,Lau:2024mqm,Salcedo:2025ezu}.
One can instead keep environments dynamical with kernels localized under the hydrodynamic derivative expansion~\cite{Haehl:2015foa,Haehl:2015uoc,Crossley:2015evo,Glorioso:2017fpd,Jensen:2017kzi}, which gives a recipe for making gauge invariance manifest~\cite{Lau:2024mqm,Abe:2026zlv}. See also Refs.~\cite{Hu:2008rga,Kaplanek:2025moq,Christodoulidis:2025vxz,Li:2026qms,Kaplanek:2026kpp} for related symmetry constraints in open systems and Refs.~\cite{Hongo:2018ant,Hongo:2019qhi,Akyuz:2023lsm} for relevant SK EFT constructions for nonequilibrium open systems.
}
Charge-changing reactions provide a mechanism for this transfer, encompassing the sphaleron-driven scenarios~\cite{Berghaus:2019whh,Berghaus:2025dqi} and their generalizations.

In Sec.~\ref{sec:invariants}, we first introduce the SK formalism and, following Ref.~\cite{Lau:2024mqm}, present a bottom-up construction of inflaton dissipation consistent with gauge invariance, clarifying the role of dynamical environmental degrees of freedom.
We then describe the environment as a thermal fluid and construct an SK effective action for warm inflation in Sec.~\ref{sec:environment-fluid}, incorporating first-order viscous and diffusive transport and slow charge-changing reactions, with terms up to Gaussian order.
In Sec.~\ref{sec:kms}, we present the dynamical KMS transformation rules and derive the thermodynamic and fluctuation--dissipation relations used in the construction.
We then specialize to a fluid with ideal stress and reaction-induced inflaton dissipation for the cosmological analysis in Sec.~\ref{sec:cosmology}.

\subsection{Schwinger--Keldysh EFT and diffeomorphism invariance}
\label{sec:invariants}

\paragraph{Schwinger--Keldysh EFT.}

\begin{figure}[t]
 \centering
 \begin{tikzpicture}[font=\small, line width=0.7pt,
   >={Stealth[length=2.3mm,width=1.6mm]}]
  \draw (0,0.55) -- (6,0.55)
   .. controls (6.55,0.55) and (6.55,-0.55) .. (6,-0.55)
   -- (0,-0.55);
  \draw[->] (2.65,0.55) -- (3.35,0.55);
  \draw[->] (3.35,-0.55) -- (2.65,-0.55);
  \fill (0,0.55) circle (1.2pt);
  \fill (0,-0.55) circle (1.2pt);
  \node at (-1.05,0) {$\widehat\rho_i$};
  \node[above=6pt] at (3,0.55) {$\Xi_+,\;\mathcal J_+$};
  \node[below=6pt] at (3,-0.55) {$\Xi_-,\;\mathcal J_-$};
  \node[above=6pt] at (0,0.55) {$t=t_i$};
  \node[above=6pt] at (6.42,0.55) {$t=t_f$};
 \end{tikzpicture}
 \caption{The Schwinger--Keldysh closed time path. Evolution runs from $t_i$ to $t_f$ on the $+$ branch and returns on the $-$ branch. The trace joins the branches at $t_f$, while the density matrix $\widehat\rho_i$ specifies the state at $t_i$.}
 \label{fig:sk-closed-time-path}
\end{figure}
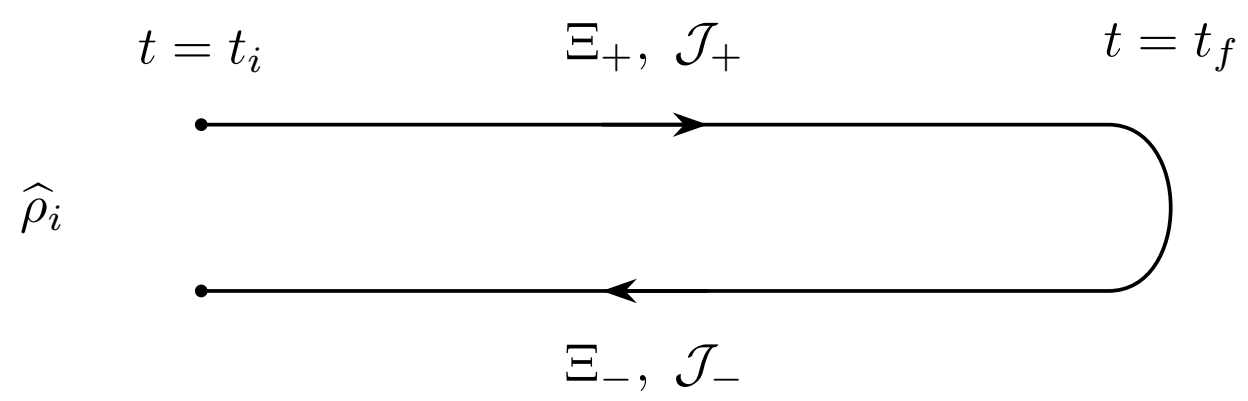

First, let us introduce the SK formalism to describe real-time evolution from an initial density operator $\widehat\rho_i$ \cite{Schwinger:1960qe,Keldysh:1964ud}.
The evolution of a density operator involves an evolution operator and its adjoint, represented by the forward and backward branches of a closed time path.
We collectively denote the microscopic fields of the system and its environment by $\Xi$.
We introduce two copies $\Xi_\pm$ of these fields, with $\Xi_+$ on the forward branch from $t_i$ to $t_f$ and $\Xi_-$ on the backward branch from $t_f$ to $t_i$, as shown in Fig.~\ref{fig:sk-closed-time-path}.
For the microscopic action $S_{\mathrm{micro}}$, the generating functional is
\begin{align}
 Z[\mathcal J_+,\mathcal J_-]
 &\equiv\operatorname{Tr}\!\left(U_{\mathcal J_+}\widehat\rho_i
 U_{\mathcal J_-}^{\dagger}\right)=\int\mathcal D\Xi_+\mathcal D\Xi_-\,
 \rho_i[\Xi_+^i,\Xi_-^i]\,
 e^{\mathrm i S_{\mathrm{micro}}[\Xi_+;\mathcal J_+]
   -\mathrm i S_{\mathrm{micro}}[\Xi_-;\mathcal J_-]},
 \label{eq:microscopic-sk-path-integral}
\end{align}
where we introduced independent sources $\mathcal J_\pm$ coupled to the fields on the respective branches. $U_{\mathcal J_\pm}$ is the evolution operator from $t_i$ to $t_f$ in the presence of $\mathcal J_\pm$, so $U_{\mathcal J_-}^{\dagger}$ generates the backward evolution.
The initial matrix element is $\rho_i[\Xi_+^i,\Xi_-^i]\equiv\langle\Xi_+^i|\widehat\rho_i|\Xi_-^i\rangle$ with $\Xi_\pm^i\equiv\Xi_\pm(t_i)$ being the initial field configurations.
The trace joins the two branches at $t_f$ and includes an integration over the final field configuration.

For an effective description on the time scales of interest, we retain the system fields together with the long-lived collective modes of the microscopic environment $\chi$ and denote the full set of retained fields by $\Psi$.
Integrating out the remaining fast modes of $\chi$ yields an SK effective action $S_{\mathrm{SK}}$ for the retained fields $\Psi_\pm$ \cite{Feynman:1963fq}.
The same generating functional can then be written as
\begin{align}
 Z[\mathcal J_+,\mathcal J_-]
 &=\int_{\rho_i}\mathcal D\Psi_+\mathcal D\Psi_-\,
 e^{\mathrm i S_{\mathrm{SK}}[\Psi_+,\Psi_-;\mathcal J_+,\mathcal J_-]},
 \label{eq:effective-sk-path-integral}
\end{align}
where the subscript $\rho_i$ denotes the initial-state weighting inherited from Eq.~\eqref{eq:microscopic-sk-path-integral}.
The two branches are joined at $t_f$.
The effective action $S_{\mathrm{SK}}$ is complex and contains couplings between the two branches that describe dissipation and fluctuations.
We define the average and response fields by
\begin{align}
 \Psi&\equiv\frac{\Psi_++\Psi_-}{2},\qquad
 \Psi_\Delta\equiv\Psi_+-\Psi_-,
 \label{eq:sk-average-response}
\end{align}
in terms of which we denote the SK effective action without sources by
\begin{align}
S_{\rm SK}[\Psi,\Psi_\Delta]\equiv S_{\rm SK}[\Psi+\tfrac{1}{2}\Psi_\Delta,\Psi-\tfrac{1}{2}\Psi_\Delta;0,0].
\end{align}
Unitarity and Hermiticity imply the SK consistency conditions \cite{Crossley:2015evo},
\begin{align}
 S_{\mathrm{SK}}[\Psi,0]&=0,\qquad
 S_{\mathrm{SK}}[\Psi,\Psi_\Delta]^*
 =-S_{\mathrm{SK}}[\Psi,-\Psi_\Delta],
 \label{eq:sk-consistency}
\end{align}
together with the positivity condition,
\begin{align}
 \operatorname{Im}S_{\mathrm{SK}}[\Psi,\Psi_\Delta]&\geq0.
 \label{eq:sk-positivity}
\end{align}
Note that odd response terms are real, whereas even ones imaginary.
In this paper, we retain first-order viscous and diffusive transport and slow charge-changing reactions.
We expand the local effective action through quadratic order in the response fields.
Unless otherwise stated, terms of order $\mathcal O(\Delta^3)$ and higher are omitted.\footnote{
We denote $n$-th and higher order terms in response variables by $\mathcal{O}(\Delta^n)$.
}
Positivity requires a positive semidefinite noise covariance and, together with thermal symmetry, fixes the sign of entropy production below.

We now apply this framework to a canonical inflaton $\phi$ coupled to gravity and microscopic environmental fields $\chi$, collectively $\Xi=(g_{\mu\nu},\phi,\chi)$, with single-copy action,
\begin{align}
 S_{\mathrm{micro}}[g,\phi,\chi]
 &=S_{g\phi}[g,\phi]+S_{\mathrm{env}}[g,\chi]
   +S_{\mathrm{int}}[g,\phi,\chi],
 \label{eq:microscopic-starting-action}\\
 S_{g\phi}[g,\phi]
 &=\int\dd^4x\sqrt{-g}\left[
 \frac{M_{\mathrm{Pl}}^2}{2}R
 -\frac12g^{\mu\nu}\partial_\mu\phi\partial_\nu\phi-V(\phi)\right],
 \label{eq:single-copy-gravity-scalar}
\end{align}
where $R$ is the Ricci scalar and $V(\phi)$ is the inflaton potential.
We leave the environmental action and interactions unspecified for later matching.
Gravitational boundary terms and gauge fixing are implicit.
Coarse-graining the fast microscopic dynamics of $\chi$, we retain its long-lived collective modes as $\mathcal H$, without yet specifying their effective description.
The fields in the effective theory are then $\Psi=(g_{\mu\nu},\phi,\mathcal H)$, and the bulk SK action is schematically of the form,
\begin{align}
 S_{\mathrm{SK}}
 &=S_{g\phi}[g_+,\phi_+]-S_{g\phi}[g_-,\phi_-]
   +S_{\mathrm{env}}^{\mathrm{SK}}[g_+,\phi_+,\mathcal H_+,g_-,\phi_-,\mathcal H_-].
 \label{eq:sk-action-decomposition}
\end{align}
The environmental term $S_{\rm env}^{\rm SK}$ describes the dynamics of the slow environmental modes $\mathcal H_\pm$ and their interactions with the inflaton and gravity, including dissipation and noise. We emphasize that $\mathcal H_\pm$ remain dynamical path-integral variables, with their initial-state distribution specified by a boundary functional. Integrating them out would generally produce a nonlocal effective action, unless we make an additional assumption on separation of scales.

{
\paragraph{Physical and noise diffeomorphisms.}
Following Ref.~\cite{Lau:2024mqm}, we require the bulk SK action with dynamical gravity to be invariant under diffeomorphisms on both branches of the contour.}
Since diffeomorphisms are local gauge symmetries, we preserve them on both SK branches in constructing the bulk effective action.
Each SK branch has an independent diffeomorphism with parameter $\varepsilon_\pm^\mu$,
\begin{align}
 \delta g_{\pm\mu\nu}&=-\mathcal L_{\varepsilon_\pm}g_{\pm\mu\nu},
 \qquad
 \delta\phi_\pm=-\mathcal L_{\varepsilon_\pm}\phi_\pm,
 \label{eq:branch-diffeomorphisms}
\end{align}
where $\mathcal L_v$ denotes the Lie derivative along an arbitrary vector field $v^\mu$.
We reparametrize the diffeomorphisms using the average and difference parameters defined by
\begin{align}
 \varepsilon^\mu&\equiv\frac{\varepsilon_+^\mu+\varepsilon_-^\mu}{2},
 \qquad
 \varepsilon_\Delta^\mu\equiv\varepsilon_+^\mu-\varepsilon_-^\mu.
 \label{eq:diffeomorphism-parameters}
\end{align}
The average parameter $\varepsilon^\mu$ generates physical diffeomorphisms, under which average and response fields are ordinary tensors:
\begin{align}
 \delta_\varepsilon g_{\mu\nu}=-\mathcal L_\varepsilon g_{\mu\nu},\qquad
 \delta_\varepsilon g_{\Delta\mu\nu}=-\mathcal L_\varepsilon g_{\Delta\mu\nu},\qquad
 \delta_\varepsilon\phi=-\mathcal L_\varepsilon\phi,\qquad
 \delta_\varepsilon\phi_\Delta=-\mathcal L_\varepsilon\phi_\Delta.
 \label{eq:physical-diffeomorphisms}
\end{align}
The difference parameter $\varepsilon^\mu_\Delta$ generates noise diffeomorphisms.
Counting $\varepsilon_\Delta^\mu$ as first order in the response-field expansion, the variations of the average fields start at second order.
At leading order, the average fields are unchanged and the response fields shift as
\begin{align}
 \delta g_{\Delta\mu\nu}&=-2\nabla_{(\mu}\varepsilon_{\Delta\nu)},
 \qquad
 \delta\phi_\Delta=-\varepsilon_\Delta^\mu\partial_\mu\phi,
 \label{eq:noise-diffeomorphism}
\end{align}
where parentheses and square brackets on indices denote normalized symmetrization and antisymmetrization, respectively, e.g., $A_{(\mu\nu)}\equiv(A_{\mu\nu}+A_{\nu\mu})/2$ and $A_{[\mu\nu]}\equiv(A_{\mu\nu}-A_{\nu\mu})/2$.

\paragraph{Diffeomorphism invariance and the St\"uckelberg trick.}
We first consider the conservative gravity--scalar sector of Eq.~\eqref{eq:sk-action-decomposition}. Expanding the difference of the two branch actions to linear order in the response fields, we obtain
\begin{align}
 S_{g\phi}[g_+,\phi_+]-S_{g\phi}[g_-,\phi_-]
 &=\int\dd^4x\sqrt{-g}\left[
 \frac12\left(T_\phi^{\mu\nu}-M_{\mathrm{Pl}}^2G_{\mathrm E}^{\mu\nu}\right)
 g_{\Delta\mu\nu}+(\Box\phi-V')\phi_\Delta\right]
 +\mathcal O(\Delta^3),
 \label{eq:conservative-response-action}
\end{align}
where $G_{\mathrm E}^{\mu\nu}\equiv R^{\mu\nu}-\tfrac12g^{\mu\nu}R$ is the Einstein tensor, $R^{\mu\nu}$ is the Ricci tensor, and primes denote derivatives with respect to $\phi$. The scalar stress tensor and its divergence are
\begin{align}
 T_\phi^{\mu\nu}&\equiv\partial^\mu\phi\partial^\nu\phi
 -g^{\mu\nu}\left[\frac12(\partial\phi)^2+V(\phi)\right],
 \label{eq:scalar-stress-definition}\\
 \nabla_\mu T_\phi^{\mu\nu}&=(\Box\phi-V')\partial^\nu\phi.
 \label{eq:scalar-stress-identity}
\end{align}
Varying $g_{\Delta\mu\nu}$ and $\phi_\Delta$ gives the Einstein equation $M_{\mathrm{Pl}}^2G_{\mathrm E}^{\mu\nu}=T_\phi^{\mu\nu}$ and the scalar equation $\Box\phi-V'=0$, respectively. Under physical diffeomorphisms, the integrand in Eq.~\eqref{eq:conservative-response-action} transforms as a scalar density. Under noise diffeomorphisms, substituting the shifts in Eq.~\eqref{eq:noise-diffeomorphism} and integrating by parts gives the bulk variation
\begin{align}
 \int\dd^4x\sqrt{-g}\,\varepsilon_{\Delta\nu}
 \left[\nabla_\mu\left(T_\phi^{\mu\nu}-M_{\mathrm{Pl}}^2G_{\mathrm E}^{\mu\nu}\right)
 -(\Box\phi-V')\partial^\nu\phi\right]=0.
 \label{eq:conservative-noise-variation}
\end{align}
The equality follows from the Bianchi identity and Eq.~\eqref{eq:scalar-stress-identity}, both of which hold off-shell. 

As a first attempt to incorporate dissipation in a bottom-up EFT, we add the following local friction term to Eq.~\eqref{eq:conservative-response-action},
\begin{align}
 -\int\dd^4x\sqrt{-g}\,\gamma_0u^\nu\nabla_\nu\phi\,\phi_\Delta,
 \label{eq:bare-scalar-drag-term}
\end{align}
where $u^\mu$ is a four-velocity and $\gamma_0$ is the friction coefficient. The term preserves physical diffeomorphism invariance when $u^\mu$ transforms as a vector. This term, however, breaks noise diffeomorphism invariance because the shift of $\phi_\Delta$ in Eq.~\eqref{eq:noise-diffeomorphism} produces a nonzero variation at this order. To restore the invariance, we introduce a St\"uckelberg field $X_\Delta^\mu$ with the transformation rules
\begin{align}
 \delta_{\varepsilon_\Delta}X_\Delta^\mu=\varepsilon_\Delta^\mu,\qquad
 \delta_\varepsilon X_\Delta^\mu=-\mathcal L_\varepsilon X_\Delta^\mu+\mathcal O(\Delta^3).
\end{align}
We can then construct the response fields
\begin{align}
 G_{\Delta\mu\nu}&\equiv g_{\Delta\mu\nu}
 +2\nabla_{(\mu}X_{\Delta\nu)}+\mathcal O(\Delta^3),
 \qquad
 \Phi_\Delta\equiv\phi_\Delta+X_\Delta^\mu\nabla_\mu\phi+\mathcal O(\Delta^3),
 \label{eq:gravity-invariants}
\end{align}
which are invariant under noise diffeomorphisms and covariant under physical diffeomorphisms to the order shown. Since higher order terms in response fields are irrelevant to our analysis of linear perturbations, we will omit $\mathcal{O}(\Delta^3)$ in the following. To make noise diffeomorphism invariance manifest, we replace $g_{\Delta\mu\nu}$ by $G_{\Delta\mu\nu}$ and $\phi_\Delta$ by $\Phi_\Delta$ in the sum of the conservative action~\eqref{eq:conservative-response-action} and the friction term~\eqref{eq:bare-scalar-drag-term}. 
The friction term in Eq.~\eqref{eq:bare-scalar-drag-term} then takes the form
\begin{align}
 -\int\dd^4x\sqrt{-g}\,\gamma_0u^\nu\nabla_\nu\phi\,\Phi_\Delta.
 \label{eq:scalar-drag-term}
\end{align}
We can thus include the friction in the SK action while keeping diffeomorphism invariance manifest.

\paragraph{Importance of long-lived environmental modes.}
The St\"uckelberg construction gives a diffeomorphism-invariant SK action at this order. The equations of motion nevertheless reveal an obstruction to dissipation. Varying the action by $\phi_\Delta$ gives the scalar equation with friction,
\begin{align}
 \Box\phi-V'=\gamma_0u^\nu\nabla_\nu\phi.
\end{align}
The $X_\Delta^\mu$ equation of motion, however, imposes the constraint
\begin{align}
 \gamma_0(u^\nu\nabla_\nu\phi)\partial_\mu\phi=0.
\end{align}
Substituting the scalar equation with friction into the stress-tensor identity~\eqref{eq:scalar-stress-identity}, we see that this constraint enforces $\nabla_\nu T_\phi^\nu{}_\mu=0$. The constraint therefore sets the energy and momentum transfer due to friction to zero. To allow this transfer while conserving total energy and momentum, we retain the long-lived environmental modes $\mathcal H$ in Eq.~\eqref{eq:sk-action-decomposition}. Their stress tensor enters the Einstein equation and accounts for the energy and momentum lost by the inflaton. In the next section, we model these modes as a fluid. See Ref.~\cite{Lau:2024mqm} for more details.

\subsection{Dynamical environment and fluid description}
\label{sec:environment-fluid}
{For warm inflation, we model the thermal environment by locally equilibrated fluids. As discussed in Sec.~\ref{sec:invariants}, we retain the long-lived environmental modes $\mathcal H$ to account for the energy and momentum transferred from the inflaton. We realize $\mathcal H$ through the local energy and momentum densities of the fluid~\cite{Crossley:2015evo}. We now develop the SK action for the dynamical environment, starting with a neutral fluid and then extending it to include conserved $U(1)$ charges and charge-changing reactions.

\subsubsection{Neutral fluid environment}
To describe the fluid dynamics, we introduce the average partner of $X_\Delta^\mu$ and identify it with the fluid map $X^\mu(\sigma)$, where $\sigma^i$ labels fluid elements and $\sigma^0$ parametrizes their worldlines. Choosing $\sigma^0$ as a thermal-time coordinate, we define the fluid velocity and thermal vector by
\begin{align}
 u^\mu(X(\sigma))&\equiv
 \frac{1}{\sqrt{-g_{\alpha\beta}(X)\partial_0X^\alpha\partial_0X^\beta}}
 \frac{\partial X^\mu}{\partial \sigma^0},
 \qquad
 \beta^\mu\equiv\frac{1}{T_0}\frac{\partial X^\mu}{\partial \sigma^0}
 =\frac{u^\mu}{T},
 \label{eq:fluid-thermal-velocity}
\end{align}
where $T_0>0$ is a constant reference temperature and $T\equiv T_0/\sqrt{-g_{\alpha\beta}(X)\partial_0X^\alpha\partial_0X^\beta}$ is the local temperature.
The fluid symmetries in these coordinates are time-independent relabelings $\sigma^i \to f^i(\sigma^j)$ and shifts of the time origin on each fluid worldline $\sigma^0 \to \sigma^0 + \lambda(\sigma^i)$ (see Ref.~\cite{Liu:2018kfw} for details).
Using the fluid map $x^\mu=X^\mu(\sigma)$, we express the fluid variables as functions of the physical spacetime coordinates $x^\mu$.
The derivative expansion requires proper frequencies and wavenumbers well below the microscopic equilibration rate $\Gamma_{\mathrm{eq}}$. We further impose dynamical KMS symmetry on the fluid EFT. In constructing the actions below, we use the thermodynamic and fluctuation--dissipation relations derived from this symmetry presented in Sec.~\ref{sec:kms}.}

We now construct the SK effective action for a dissipative neutral fluid coupled to the inflaton. {The action incorporates scalar friction, first-order viscous corrections, and their associated noise terms.\footnote{Dissipative scalar--radiation systems with these ingredients were studied in cosmology \cite{Bastero-Gil:2014jsa}. In particular, shear viscosity can damp radiation fluctuations and thereby modify the curvature power spectrum in warm inflation \cite{Bastero-Gil:2011rva}.}
The zeroth-order contribution to the fluid stress tensor takes the ideal-fluid form,
\begin{align}
 T_{\mathrm h}^{\mu\nu}&=\rho u^\mu u^\nu+P\Delta^{\mu\nu},\qquad \Delta^{\mu\nu}\equiv g^{\mu\nu}+u^\mu u^\nu,\qquad u_\mu u^\mu=-1.
 \label{eq:fluid-stress}
\end{align}
For the neutral fluid, the equation of state is specified by $P(T)$, with entropy density $s(T)=\partial P/\partial T$ and energy density $\rho(T)=Ts(T)-P(T)$.
We work in the Landau frame, where  a dissipative stress is transverse to $u^\mu$. The expansion and shear entering the first-order viscous stress are defined by
\begin{align}
 \theta&\equiv\nabla_\mu u^\mu,\qquad
 \Sigma^{\mu\nu}\equiv\mathcal P^{\mu\nu,\alpha\beta}\nabla_\alpha u_\beta,\qquad
 \mathcal P^{\mu\nu,\alpha\beta}\equiv\Delta^{\mu(\alpha}\Delta^{\beta)\nu}
 -\frac13\Delta^{\mu\nu}\Delta^{\alpha\beta}.
 \label{eq:viscous-kinematics}
\end{align}
For inflaton-friction and viscous sectors, the scalar friction is $\gamma_0u^\mu\nabla_\mu\phi$ and the dissipative stress is
\begin{align}
 \Pi^{\mu\nu}
 &=-2\eta\Sigma^{\mu\nu}-\zeta\theta\Delta^{\mu\nu}.
 \label{eq:drag-viscous-response}
\end{align}
The dimensions of $\gamma_0(T)$ and the viscosities $\eta(T),\, \zeta(T)$ are one and three, respectively, and we treat them as local constitutive functions of $T$.\footnote{
{$\phi$-dependence of $\gamma_0,\eta,\zeta$ is compatible with the fluid symmetry, but we ignore it under an assumption analogous to the slowroll conditions. The same remark applies also to other parameters introduced in the following.}}
The fluid-sector contribution to the SK action, including the viscous response and its associated noise, is
\begin{align}
    S_{\mathrm h}
    =\int \dd^4x\sqrt{-g}\Bigg[
    &\frac12\mathcal{T}_{\mathrm h}^{\mu\nu}
    G_{\Delta\mu\nu}+\frac{\i C_\eta}{8}
    \mathcal P^{\mu\nu,\alpha\beta}   G_{\Delta\mu\nu}G_{\Delta\alpha\beta}+\frac{\i C_\zeta}{8}
    \Delta^{\mu\nu}\Delta^{\alpha\beta}
    G_{\Delta\mu\nu}G_{\Delta\alpha\beta}
    \Bigg],
    \label{eq:fluid-sk-action}
\end{align}
where $\mathcal{T}_{\mathrm h}^{\mu\nu}=T_{\mathrm h}^{\mu\nu}+\Pi^{\mu\nu}$.
The scalar-friction, shear, and bulk noise coefficients are denoted by $N_0$, $C_\eta$, and $C_\zeta$, respectively.
These coefficients are a priori independent of the corresponding dissipative coefficients. Imposing dynamical KMS symmetry gives}
\begin{align}
 N_0
 &=2T\gamma_0,
 \qquad
 C_\eta=4T\eta,
 \qquad
 C_\zeta=2T\zeta,
 \label{eq:neutral-noise-kms}
\end{align}
while SK positivity \eqref{eq:sk-positivity} requires $\gamma_0,\eta,\zeta\geq0$. Combining the gravity, inflaton and fluid contributions, the action is
\begin{align}
 S_{\mathrm{neutral}}
 ={}&\int\dd^4x\sqrt{-g}\Bigg\{
 \frac12\left(T_\phi^{\mu\nu}+T_{\mathrm h}^{\mu\nu}-M_{\mathrm{Pl}}^2G_{\mathrm E}^{\mu\nu}\right)G_{\Delta\mu\nu}
 \nonumber\\
 &\quad
 +\left(\Box\phi-V'-\gamma_0u^\mu\nabla_\mu\phi\right)\Phi_\Delta
 -\eta\Sigma^{\mu\nu}G_{\Delta\mu\nu}
 -\frac{\zeta}{2}\theta\Delta^{\mu\nu}G_{\Delta\mu\nu}
 \nonumber\\
 &\quad
 +\mathrm iT\gamma_0\Phi_\Delta^2
 +\frac{\mathrm iT\eta}{2}\mathcal P^{\mu\nu,\alpha\beta}G_{\Delta\mu\nu}G_{\Delta\alpha\beta}
 +\frac{\mathrm iT\zeta}{4}\Delta^{\mu\nu}\Delta^{\alpha\beta}G_{\Delta\mu\nu}G_{\Delta\alpha\beta}
 \Bigg\}.
 \label{eq:neutral-fluid-action}
\end{align}
We thus constructed an effective action describing warm inflation with a neutral fluid environment.

\subsubsection{Fluid environment with conserved \texorpdfstring{$U(1)$}{U(1)} charges}

We now extend the neutral-fluid construction to a fluid carrying $m$ conserved $U(1)$ charges, labeled by $a,b=1,\ldots,m$.
Repeated charge indices are summed within each term, irrespective of index position, unless otherwise stated.
To encode the Ward identities of the slow internal charges, we introduce a nondynamical current probe $A_{a\mu}$ and a hydrodynamic phase $\psi_a$ for each charge. The average and difference of the independent branch parameters $\varphi_{a,\pm}$ are $\varphi_a\equiv(\varphi_{a,+}+\varphi_{a,-})/2$ and $\varphi_{a\Delta}\equiv\varphi_{a,+}-\varphi_{a,-}$. The physical charge-source transformation generated by $\varphi_a$ acts only on the average source and phase, and the noise charge-source transformation generated by $\varphi_{a\Delta}$ acts only on their response counterparts. The nonzero variations are
\begin{align}
 \begin{aligned}
 \delta_{\varphi} A_{a\mu}&=\partial_\mu\varphi_a,
 &\qquad \delta_{\varphi}\psi_a&=-\varphi_a,\\
 \delta_{\varphi_\Delta} A_{a\Delta\mu}&=\partial_\mu\varphi_{a\Delta},
 &\qquad \delta_{\varphi_\Delta}\psi_{a\Delta}&=-\varphi_{a\Delta}.
 \end{aligned}
 \label{eq:source-transformations}
\end{align}
The inflaton is neutral under both transformations. In the physical sector, the invariant one-form and the corresponding chemical potential are
\begin{align}
 B_{a\mu}&\equiv A_{a\mu}+\partial_\mu\psi_a,\qquad \mu_a\equiv u^\mu B_{a\mu}.
 \label{eq:chemical-one-form}
\end{align}
Completing the response with the St\"uckelberg field $X_\Delta^\mu$ gives
\begin{align}
 B_{a\Delta\mu}&\equiv A_{a\Delta\mu}+\partial_\mu\psi_{a\Delta}+\mathcal L_{X_\Delta}B_{a\mu}.
 \label{eq:charge-response-one-form}
\end{align}
Like the metric and scalar responses in Eq.~\eqref{eq:gravity-invariants}, $B_{a\Delta\mu}$ is invariant under noise diffeomorphisms at $\order{\Delta^2}$.
For a normal fluid, we impose the chemical shift symmetry $\psi_a\to\psi_a+\lambda_a(\sigma^i)$, under which $B_{a\mu}\to B_{a\mu}+\partial_\mu\lambda_a$ \cite{Dubovsky:2011sj,Glorioso:2017fpd}. Since $u^\mu\partial_\mu\lambda_a=0$, the ideal constitutive relations depend on $B_{a\mu}$ only through $\mu_a$.
We denote the current of charge $a$ by $j_a^\mu$.
The coupling $ j_a^\mu B_{a\Delta\mu}$ gives the charge equations, with
\begin{align}
 j_a^\mu&=n_a u^\mu+d_a^\mu,\qquad u_\mu d_a^\mu=0,
 \label{eq:charge-current}
\end{align}
where $n_a$ is the charge density in the fluid rest frame.
For the charged fluid, the equation of state is extended to $P(T,\mu_a)$, with $n_a=\partial P/\partial\mu_a$ and $\rho=Ts+\mu_a n_a-P$.
The friction and viscosity coefficients can also depend on the chemical potentials.
We parameterize the diffusive current $d_a^\mu$ by a conductivity matrix $\sigma_{ab}$ of mass dimension one:
\begin{align}
 d_a^\mu&=-T\sigma_{ab}\Delta^{\mu\nu} \mathcal L_\beta B_{b\nu}.
 \label{eq:diffusion}
\end{align}
Cartan's identity expresses the thermal derivative in terms of the field strength:
\begin{align}
 \mathcal L_\beta B_{b\nu}&=\partial_\nu(\mu_b/T)+\beta^\lambda F_{b\lambda\nu},
 \qquad F_{b\mu\nu}\equiv\partial_\mu A_{b\nu}-\partial_\nu A_{b\mu}.
 \label{eq:thermal-charge-derivative}
\end{align}

Dynamical KMS symmetry fixes the diffusion noise covariance to $2T\sigma_{(ab)}\Delta^{\mu\nu}$, while SK positivity requires $\sigma_{(ab)}\succeq 0$, where $\succeq0$ denotes positive semidefiniteness. With these constraints imposed, we combine the charge terms with the neutral-fluid action in Eq.~\eqref{eq:neutral-fluid-action} to obtain
\begin{align}
 S_{\mathrm{charged}}
 &=\int\dd^4x\sqrt{-g}\Bigg\{
 \frac12\left(T_\phi^{\mu\nu}+T_{\mathrm h}^{\mu\nu}-M_{\mathrm{Pl}}^2G_{\mathrm E}^{\mu\nu}\right)G_{\Delta\mu\nu}+\left(\Box\phi-V'-\gamma_0u^\mu\nabla_\mu\phi\right)\Phi_\Delta
 \nonumber\\
 &
 -\eta\Sigma^{\mu\nu}G_{\Delta\mu\nu}
 -\frac{\zeta}{2}\theta\Delta^{\mu\nu}G_{\Delta\mu\nu}
+n_a u^\mu B_{a\Delta\mu}
 -T\sigma_{ab}\Delta^{\mu\nu}\left(\mathcal L_\beta B_{b\nu}\right)B_{a\Delta\mu}
 \nonumber\\
 &
 +\mathrm iT\gamma_0\Phi_\Delta^2
 +\frac{\mathrm iT\eta}{2}\mathcal P^{\mu\nu,\alpha\beta}G_{\Delta\mu\nu}G_{\Delta\alpha\beta}
 +\frac{\mathrm iT\zeta}{4}\Delta^{\mu\nu}\Delta^{\alpha\beta}G_{\Delta\mu\nu}G_{\Delta\alpha\beta}
 +\mathrm iT\sigma_{(ab)}\Delta^{\mu\nu}B_{a\Delta\mu}B_{b\Delta\nu}
 \Bigg\},
 \label{eq:charged-fluid-action}
\end{align}
{where the EFT coefficients $\eta,\zeta,n_a,\sigma_{ab}$ are functions of $T,\mu_a$.
While the above source formulation readily extends to dynamical gauge fields,\footnote{{
We can make gauge fields dynamical simply by adding the gauge-field action with opposite signs on the two SK branches. For related SK effective theories of dynamical gauge fields coupled to media on fixed spacetime backgrounds, see Refs.~\cite{Landry:2022nog,Salcedo:2024nex,Abe:2026zlv}. The resulting theory couples gauge fields to the charged fluid and gravity, with potential applications to the evolution of primordial magnetic fields and the gravitational waves they source.
}
} in the warm inflation models studied below, we focus on the case with (approximate) global symmetries. For this, it is convenient to set the external probes to zero, $A_{a\mu}=A_{a\Delta\mu}=0$, and so the charge-sector one-forms reduce to
\begin{align}
 B_{a\mu}&=\partial_\mu\psi_a,\qquad
 B_{a\Delta\mu}=\partial_\mu\left(\psi_{a\Delta}+X_\Delta^\nu\partial_\nu\psi_a\right).
 \label{eq:zero-source-response}
\end{align}}

\subsubsection{Charge-changing reactions in the fluid environment}

We finally include $N_{\mathrm{reac}}$ charge-changing reactions whose relaxation rates are smaller than the microscopic equilibration rate of the fluid. We label the reactions by $r,s=1,\ldots,N_{\mathrm{reac}}$, with repeated reaction indices summed irrespective of index position unless stated otherwise. When coupled to the inflaton, these processes transfer energy between the scalar and the fluid and contribute to scalar friction. We reserve $\gamma_0$ for friction from mechanisms not included in these reaction channels. Each reaction channel is specified by the dimensionless inputs $c_r$ and $\nu_{ra}$. A positive event changes charge $a$ by $-\nu_{ra}$ and contributes a scalar force $-c_r/f$, where $f$ sets the inflaton coupling scale. In the sphaleron-heating example, $f$ corresponds to the scale of the axion-like interaction in Eq.~\eqref{eq:microscopic-axion-coupling}.

At zero external sources, an infinitesimal advance $\dd\xi_r$ of reaction $r$ transfers energy $\dd E=(c_r/f)u^\mu\nabla_\mu\phi\,\dd\xi_r$ to the fluid in its local rest frame and changes its charges by $\dd Q_a=-\nu_{ra}\dd \xi_r$.
Hence, from the first law at fixed fluid volume, variation of the fluid entropy $S_{\mathrm{reac}}$ reads
\begin{align}
 T\,\dd S_{\mathrm{reac}}&=\dd E-\mu_a\,\dd Q_a
 =\left(\frac{c_r}{f}u^\mu\nabla_\mu\phi+\nu_{ra}\mu_a\right)\dd\xi_r,
 \label{eq:reaction-affinity-first-law}
\end{align}
where no sum over $r$ is taken.
The coefficient of $\dd\xi_r$ is the thermodynamic affinity of reaction $r$.

The hydrodynamic phases $\psi_a$ and their response fields $\psi_{a\Delta}$ remain dynamical, with $\mu_a=u^\mu\partial_\mu\psi_a$. To encode the energy and charge transfers of each reaction in the SK action, we introduce a covariant reaction response $\mathcal A_{r\Delta}$. The thermodynamic affinity $\mathcal A_r$ identified in Eq.~\eqref{eq:reaction-affinity-first-law} and the reaction response are given by
\begin{align}
 \mathcal A_r
 &\equiv\frac{c_r}{f}u^\mu\nabla_\mu\phi+\nu_{ra}\mu_a,
\qquad
 \mathcal A_{r\Delta}
 \equiv\frac{c_r}{f}\Phi_\Delta
 +\nu_{ra}\left(\psi_{a\Delta}
 +X_\Delta^\mu\partial_\mu\psi_a\right).
 \label{eq:zero-source-affinity}
\end{align}
Scalar and chemical responses thus enter the same reaction combination $\mathcal A_{r\Delta}$.\footnote{
For nonzero sources, the charge violation can be implemented in a source-covariant form by introducing a prescribed breaking phase $\vartheta_{r,\pm}$ transforming as
$\vartheta_{r,\pm}\to\vartheta_{r,\pm}+\nu_{ra}\varphi_{a,\pm}$.
The combination $(c_r/f)\phi+\nu_{ra}\psi_a+\vartheta_r$ is then invariant under the charge-source transformations. Its fluid derivative gives the source-covariant affinity,
$\mathcal A_r=u^\mu\partial_\mu[(c_r/f)\phi+\nu_{ra}\psi_a+\vartheta_r]$, as well as the corresponding response
$\mathcal A_{r\Delta}=(c_r/f)\Phi_\Delta
+\nu_{ra}\psi_{a\Delta}+\vartheta_{r\Delta}
+X_\Delta^\mu\partial_\mu(\nu_{ra}\psi_a+\vartheta_r)$.
This spurion construction does not impose exact charge conservation at fixed physical sources~\cite{Landry:2020tbh,Hongo:2024brb}.
}
We couple this combination to the reaction currents in the SK action to generate the corresponding scalar force and charge changes.

\paragraph{The dissipative Einstein--scalar--charged-fluid action.}
We now combine the ingredients introduced above to write the SK action for the coupled system. Scalar friction, bulk viscosity and reactions can mix because their forces and responses are scalars under spatial rotations in the local rest frame. We collect the corresponding forces and responses as
\begin{align}
 \boldsymbol F&\equiv\bigl(u^\mu\nabla_\mu\phi,\theta,\{\mathcal A_r\}\bigr)^{\mathsf T},
 \qquad
 \boldsymbol q_\Delta\equiv\bigl(\Phi_\Delta,\Delta^{\mu\nu}G_{\Delta\mu\nu}/2,\{\mathcal A_{r\Delta}\}\bigr)^{\mathsf T}.
 \label{eq:scalar-transport-basis}
\end{align}
We denote the scalar transport and noise covariance matrices by $\mathsf M$ and $\mathsf K=\mathsf K^{\mathsf T}$. Writing the charge current from Eq.~\eqref{eq:charged-fluid-action} as $j_a^\mu\equiv n_a u^\mu-T\sigma_{ab}\Delta^{\mu\nu}\mathcal L_\beta B_{b\nu}$, we obtain
\begin{align}
 S_{\mathrm{gen}}&=\int\dd^4x\sqrt{-g}\Bigg\{
 \frac12\left(T_\phi^{\mu\nu}+T_{\mathrm h}^{\mu\nu}
 -M_{\mathrm{Pl}}^2G_{\mathrm E}^{\mu\nu}\right)G_{\Delta\mu\nu}
 +(\Box\phi-V')\Phi_\Delta+j_a^\mu B_{a\Delta\mu}
 -\boldsymbol q_\Delta^{\mathsf T}\mathsf M\boldsymbol F
 \nonumber\\
 &\quad+\frac{\mathrm i}{2}\boldsymbol q_\Delta^{\mathsf T}\mathsf K\boldsymbol q_\Delta
 -\eta\Sigma^{\mu\nu}G_{\Delta\mu\nu}
 +\frac{\mathrm i}{8}C_\eta\mathcal P^{\mu\nu,\alpha\beta}
 G_{\Delta\mu\nu}G_{\Delta\alpha\beta}
 +\mathrm iT\sigma_{(ab)}\Delta^{\mu\nu}
 B_{a\Delta\mu}B_{b\Delta\nu}
 \Bigg\}.
 \label{eq:general-dissipative-action}
\end{align}
This action gives the general form of our SK EFT for warm inflation with a thermal fluid environment, allowing for charge-changing reactions such as sphaleron transitions.

In Sec.~\ref{sec:reactive-model}, we specialize this action to the model used in our cosmological analysis. To facilitate this specialization, we label the scalar-friction, bulk-viscous and reaction channels by $0$, $\theta$ and $r$, respectively, and write the diagonal blocks of $\mathsf M$ and $\mathsf K$ as
\begin{align}
 \mathsf M_{00}&=\gamma_0,\qquad
 \mathsf M_{\theta\theta}=\zeta,\qquad
 \mathsf M_{rs}=\frac{L_{rs}}{T},
 \nonumber\\
 \mathsf K_{00}&=N_0,\qquad
 \mathsf K_{\theta\theta}=C_\zeta,\qquad
 \mathsf K_{rs}=N_{rs}.
\end{align}
For the friction and viscous noise coefficients, we use the dynamical KMS relations given above, $N_0=2T\gamma_0$, $C_\zeta=2T\zeta$ and $C_\eta=4T\eta$. We take the reaction transport coefficients $L_{rs}$ to be local functions of $T$ and the dimensionless affinities $\{\mathcal A_r/T\}$, and denote the reaction-noise covariance matrix by $N=(N_{rs})$. In the dissipative response term, $\mathcal A_{r\Delta}$ enters through $-\mathcal A_{r\Delta}J_r$, where the net reaction rate per unit physical volume is
\begin{align}
 J_r&=(\mathsf M\boldsymbol F)_r
 =\mathsf M_{r0}u^\mu\nabla_\mu\phi
 +\mathsf M_{r\theta}\theta
 +\frac{L_{rs}}{T}\mathcal A_s.
 \label{eq:unmixed-transport-matrices}
\end{align}

We retain the first-order viscous and diffusive operators introduced above. Magnetic-field- and vorticity-dependent transport, other anomalous contributions and higher-gradient corrections lie outside the present operator basis.

\subsection{Dynamical KMS symmetry and thermodynamics}
\label{sec:kms}
\label{sec:thermodynamics}

In Sec.~\ref{sec:environment-fluid}, we employed dynamical KMS symmetry as an input to the construction of the fluid action. We now give the transformation rules, derive the thermodynamic relations, and obtain the fluctuation--dissipation and Onsager--Casimir constraints on the full transport and noise matrices, including the reaction and mixed blocks. These relations, combined with SK positivity, imply nonnegative entropy production at this order.

\paragraph{Dynamical KMS transformation.}
The dynamical KMS transformation combines an involutive antiunitary operation $\Theta$ containing time reversal with imaginary thermal shifts of the response fields \cite{Glorioso:2017fpd,Liu:2018kfw}.
We leave the choice of T, PT or CPT open at this stage.
We choose $\Theta$ consistently with the potential, pressure, sources and reaction directions at fixed microscopic couplings; this fixes the affinity parities.
Tildes denote fields after the dynamical KMS transformation. The average fields transform under $\Theta$, including its action on coordinates, tensor indices and intrinsic parities. With $\widetilde\beta^\mu=\widetilde u^\mu/\widetilde T$ and future-directed $\widetilde u^\mu$, the  covariant response fields transform as
\begin{align}
 \widetilde{G}_{\Delta\mu\nu}
 &=\Theta[G_{\Delta\mu\nu}]-\mathrm i\mathcal L_{\widetilde\beta}\widetilde g_{\mu\nu},
 \qquad
 \widetilde\Phi_\Delta=\Theta[\Phi_\Delta]-\mathrm i\mathcal L_{\widetilde\beta}\widetilde\phi,
 \qquad
 \widetilde B_{a\Delta\mu}=\Theta[B_{a\Delta\mu}]-\mathrm i\mathcal L_{\widetilde\beta}\widetilde B_{a\mu}.
 \label{eq:parent-kms-map}
\end{align}
At vanishing external probes, the Lie derivatives entering the thermal shifts are
\begin{align}
 \mathcal L_\beta g_{\mu\nu}&=2\nabla_{(\mu}\beta_{\nu)},\qquad
 \mathcal L_\beta B_{a\mu}=\partial_\mu(\mu_a/T),\qquad
 \mathcal L_\beta\phi=\frac{u^\mu\partial_\mu\phi}{T}.
 \label{eq:kms-thermal-directions}
\end{align}
Using $\beta^\mu\partial_\mu\psi_a=\mu_a/T$, we obtain the transformation of the reaction response,
\begin{align}
 \widetilde{\mathcal A}_{r\Delta}&=\Theta[\mathcal A_{r\Delta}]
 -\mathrm i\frac{\widetilde{\mathcal A}_r}{\widetilde T}.
 \label{eq:reaction-kms}
\end{align}
The dynamical KMS map defines a $\mathbb{Z}_2$ transformation, since $\Theta^2=1$ and the thermal shifts cancel when the map is applied twice. Requiring the SK action to be invariant under this map up to boundary terms gives the constraints below.

\paragraph{Ideal-fluid thermodynamics from dynamical KMS.}
Dynamical KMS invariance encodes local thermal equilibrium in the SK action. In the ideal-fluid sector, invariance requires the response-independent part of the variation to be a total derivative. Matching the term proportional to the fluid expansion fixes the ideal-order boundary current to $P\beta^\mu$, giving the off-shell identity
\begin{align}
 \nabla_\mu(P\beta^\mu)&=T_{\mathrm h}^{\mu\nu}\nabla_\mu\beta_\nu+n_a u^\mu\partial_\mu(\mu_a/T).
 \label{eq:thermal-boundary}
\end{align}
Substituting Eq.~\eqref{eq:fluid-stress} and matching the coefficients of the independent gradients $u^\mu\partial_\mu T$ and $u^\mu\partial_\mu\mu_a$, we obtain the thermodynamic identities
\begin{align}
 s&=\frac{\partial P}{\partial T},\qquad n_a=\frac{\partial P}{\partial\mu_a},\qquad \rho=Ts+\mu_a n_a-P.
 \label{eq:thermodynamics}
\end{align}
The enthalpy density $h{\equiv}\rho{+}P$ is thus $h{=}Ts{+}\mu_a n_a$. They also give the thermodynamic differentials
\begin{align}
 \dd P&=s\,\dd T+n_a\,\dd\mu_a,\qquad \dd\rho=T\,\dd s+\mu_a\,\dd n_a.
 \label{eq:thermodynamic-differentials}
\end{align}
The pressure is a matching input determining the densities and susceptibilities through the relations.

\paragraph{Dissipation, noise and Onsager--Casimir relations.}
Having established dynamical KMS invariance of the ideal-fluid sector up to a boundary term, we now turn to the dissipative part of the action, where the symmetry imposes additional constraints on the transport and noise coefficients.
Under the dynamical KMS transformation, the response vector in Eq.~\eqref{eq:scalar-transport-basis} obeys
\begin{align}
 \widetilde{\boldsymbol q}_\Delta
 &=\Theta[\boldsymbol q_\Delta]
 -\mathrm i\frac{\widetilde{\boldsymbol F}}{\widetilde T},
\end{align}
where the second term is the thermal shift of the response fields. For the bulk component, we use $\Delta^{\mu\nu}\mathcal L_\beta g_{\mu\nu}/2=\theta/T$. A tilded coefficient is evaluated on transformed data and pulled back in coordinates and both response indices, including intrinsic parities.

Requiring the general SK action in Eq.~\eqref{eq:general-dissipative-action} to be invariant under the dynamical KMS transformation, up to boundary terms and through quadratic order in the response fields, we obtain
\begin{align}
 \widetilde{\mathsf M}&=\mathsf M^{\mathsf T},\qquad
 \mathsf K=2T\mathsf M_{\mathrm{sym}},\qquad
 C_\eta=4T\eta,\qquad
 \mathsf M_{\mathrm{sym}}\equiv\frac12(\mathsf M+\mathsf M^{\mathsf T}).
 \label{eq:general-transport-kms}
\end{align}
The shear-viscosity coefficient is even under this transformation, $\widetilde\eta=\eta$. For the charge-diffusion sector, dynamical KMS symmetry fixes the associated Gaussian noise covariance to $2T\sigma_{(ab)}\Delta^{\mu\nu}$. The SK positivity $ \operatorname{Im}S_{\mathrm{SK}}\geq0$ further requires $\mathsf M_{\mathrm{sym}}\succeq0$, $\eta\geq0$ and $\sigma_{(ab)}\succeq0$ on the space of nonredundant response combinations.

The reaction and charge-diffusion sectors obey the Onsager--Casimir relations
\begin{align}
 \widetilde L_{rs}&=L_{sr},\qquad
 \widetilde N_{rs}=N_{rs},\qquad
 \widetilde\sigma_{ab}=\sigma_{ba}.
 \label{eq:constitutive-casimir}
\end{align}
In matrix notation, this is $\widetilde L=L^{\mathsf T}$ and $\widetilde N=N$. The latter relation cancels the terms quadratic in the response fields in the KMS variation. Matching the terms proportional to $\mathcal A_{r\Delta}\mathcal A_s$ then gives $\widetilde L+L=\widetilde N$ and hence the fluctuation--dissipation relation
\begin{align}\label{eq:reaction-fdt}
 N=L+L^{\mathsf T},\qquad N_{rs}=2L_{(rs)}.
\end{align}
The remaining response-independent term is proportional to $\mathcal A_r L_{[rs]}\mathcal A_s$ and vanishes by antisymmetry. For the finite-affinity backgrounds below, we use a Gaussian reaction model and impose the reaction-block relations in Eq.~\eqref{eq:general-transport-kms} on $L_{rs}$ and $N_{rs}$ at the local temperature and affinities.

\paragraph{Entropy current from dynamical KMS and positivity.}

To check the thermodynamic consistency of the coupled system, we construct its entropy current and verify that its divergence is nonnegative. Dynamical KMS symmetry identifies an entropy current, while SK positivity controls its divergence~\cite{Glorioso:2016gsa,Glorioso:2017fpd}. We apply this construction to the general dissipative action in Eq.~\eqref{eq:general-dissipative-action}. The contributions from charge diffusion, scalar transport and shear viscosity cancel against their corresponding noise terms through the fluctuation--dissipation relations. 

Let $\Psi$ and $\Psi_\Delta$ collectively denote the average and response fields, respectively. 
{Expressing the transformed action in the original coordinates and using Eq.~\eqref{eq:thermal-boundary} together with $\nabla_\mu T_\phi^{\mu\nu}=(\Box\phi-V')\nabla^\nu\phi$ and the Bianchi identity $\nabla_\mu G_{\mathrm E}^{\mu\nu}=0$, we obtain}
\begin{align}
 S_{\mathrm{gen}}[\widetilde\Psi,\widetilde\Psi_\Delta]
 -S_{\mathrm{gen}}[\Psi,\Psi_\Delta]
 &=\mathrm i\int\dd^4x\sqrt{-g}\,
 \nabla_\mu\left[
 P\beta^\mu+
 \left(T_\phi^{\mu\nu}
 -M_{\mathrm{Pl}}^2G_{\mathrm E}^{\mu\nu}\right)\beta_\nu
 \right].
\end{align}
To obtain the entropy current, we subtract from this boundary current the full metric-response coefficient contracted with $\beta_\nu$ and the chemical work. The scalar and Einstein terms cancel, leaving
\begin{align}
 S^\mu
 &=P\beta^\mu-T_{\mathrm h}^{\mu\nu}\beta_\nu
 -\frac{\mu_a}{T}j_a^\mu
 =su^\mu-\frac{\mu_a}{T}d_a^\mu.
\end{align}
Viscosity drops out of the current because $u_\nu\Pi^{\mu\nu}=0$, but enters its divergence.

Taking the divergence of the entropy current and using the scalar equation of motion, the energy--momentum balance and the charge-balance equations following from the general SK action in Eq.~\eqref{eq:general-dissipative-action}, we obtain the entropy-production rate with the full scalar transport matrix $\mathsf M$ retained:
\begin{align}
 \nabla_\mu S^\mu
 &=\frac1T\boldsymbol F^{\mathsf T}
 \mathsf M_{\mathrm{sym}}\boldsymbol F
 +\frac{2\eta}{T}\Sigma_{\mu\nu}\Sigma^{\mu\nu}
 +T\sigma_{(ab)}\Delta^{\mu\nu}
 \partial_\mu\qty(\frac{\mu_a}{T})
 \partial_\nu\qty(\frac{\mu_b}{T})
 \geq0.
 \label{eq:entropy-production}
\end{align}
The inequality follows from the SK-positivity conditions above. In particular, the antisymmetric part of $\mathsf M$ does not contribute to entropy production.

\subsection{Reactive-fluid specialization for warm inflation}
\label{sec:reactive-model}

We now specialize the general dissipative SK action and its dynamical KMS constraints to a warm-inflation setup motivated by Ref.~\cite{Berghaus:2019whh}. The charge-changing reactions introduced above account for all inflaton dissipation, and we take the stress of the charged normal fluid to be ideal.\footnote{Shear viscosity and its associated stress fluctuations can affect tensor perturbations, as studied in Ref.~\cite{Klose:2022rxh}.} We set the direct scalar-friction and bulk-viscous blocks of $\mathsf M$ and $\mathsf K$, together with all cross-couplings involving these blocks, to zero. The scalar and shear transport and noise coefficients are then
\begin{align}
 \mathsf M&=\operatorname{diag}\left(0,0,\frac{L}{T}\right),\qquad
 \mathsf K=\operatorname{diag}(0,0,N),\qquad
 \eta=0,\qquad C_\eta=0.
 \label{eq:reactive-ideal-specialization}
\end{align}
The reaction blocks $L/T$ and $N$ may have off-diagonal entries, allowing transport couplings and noise correlations between different reaction channels. Charge diffusion is retained through the conductivity matrix $\sigma_{ab}$.
With this choice of $\mathsf M$, the net reaction rate per unit physical volume reduces to
\begin{align}
 J_r&=\frac{1}{T}L_{rs}\mathcal A_s.
 \label{eq:reaction-current}
\end{align}
For the perturbation analysis in Sec.~\ref{sec:second-variation}, we linearize this current about the background,
\begin{align}
 \delta J_r&=J_{r,T}\,\delta T+J_{r,\mathcal A_s}\,\delta\mathcal A_s,
\end{align}
where $J_{r,T}\equiv(\partial J_r/\partial T)_{\{\mathcal A_\ell\}}$ and $J_{r,\mathcal A_s}\equiv(\partial J_r/\partial\mathcal A_s)_T$. At finite affinity, the dependence of $L$ on the dimensionless affinities can contribute to the current Jacobian through the second term in
\begin{align}
 J_{r,\mathcal A_s}
 &=\frac{L_{rs}}{T}
 +\frac{\mathcal A_\ell}{T^2}
 \frac{\partial L_{r\ell}}{\partial(\mathcal A_s/T)}.
 \label{eq:differential-current}
\end{align}

\paragraph{SK action and deterministic equations.}

To derive the coupled equations for the inflaton, reactive fluid and gravity, we substitute the transport and noise coefficients in Eq.~\eqref{eq:reactive-ideal-specialization} into the general SK action in Eq.~\eqref{eq:general-dissipative-action}. The resulting action is
\begin{align}
 S_{\mathrm{SK}}&=\int\dd^4x\sqrt{-g}\Bigg\{
 \frac12\left(T_\phi^{\mu\nu}+T_{\mathrm h}^{\mu\nu}
 -M_{\mathrm{Pl}}^2G_{\mathrm E}^{\mu\nu}\right)G_{\Delta\mu\nu}
 +(\Box\phi-V')\Phi_\Delta
 +j_a^\mu B_{a\Delta\mu}
 \nonumber\\
 &\quad-J_r\mathcal A_{r\Delta}
 +\frac{\mathrm i}{2}N_{rs}\mathcal A_{r\Delta}\mathcal A_{s\Delta}
 +\mathrm iT\sigma_{(ab)}\Delta^{\mu\nu}
 B_{a\Delta\mu}B_{b\Delta\nu}
 \Bigg\}.
 \label{eq:parent-action}
\end{align}
The fluid stress tensor is ideal, and there are no direct scalar-friction or viscous terms. Instead, the reaction term $-J_r\mathcal A_{r\Delta}$ mediates energy and charge transfer, while the conductivity term describes charge diffusion and its associated noise.

Varying this action with respect to the response fields gives the Einstein, scalar and charge equations. At zero external sources, they read
\begin{align}
 M_{\mathrm{Pl}}^2G_{\mathrm E}^{\mu\nu}
 &=T_\phi^{\mu\nu}+T_{\mathrm h}^{\mu\nu},
 \label{eq:einstein-balance}\\
 0&=\Box\phi-V'-\frac{c_r}{f}J_r,
 \label{eq:scalar-balance}\\
 \nabla_\mu j_a^\mu&=-\nu_{ra}J_r.
 \label{eq:scalar-charge-balances}
\end{align}
The Bianchi identity and the scalar equation then imply the fluid-energy balance,
\begin{align}
 \nabla_\mu T_{\mathrm h}^{\mu\nu}
 &=-\frac{c_r}{f}J_r\partial^\nu\phi.
 \label{eq:bath-balance}
\end{align}
Thus, the reaction currents consistently describe the transfer of energy between the inflaton and the fluid, together with the corresponding changes in the fluid charges.

With $\gamma_0=0$, charge-changing reactions provide the dissipative coupling between the inflaton and the fluid. The inflaton velocity drives the reaction currents through the affinity $\mathcal A_r=(c_r/f)u^\mu\nabla_\mu\phi+\nu_{ra}\mu_a$, and the currents in turn exert the force $-c_rJ_r/f$ on the inflaton. Substituting the affinity into Eq.~\eqref{eq:reaction-current} gives
\begin{align}
 \frac{c_r}{f}J_r
 &=\gamma_{\mathrm{reac}}u^\mu\nabla_\mu\phi
 +\frac{1}{Tf}c_rL_{rs}\nu_{sa}\mu_a,\qquad
 \gamma_{\mathrm{reac}}\equiv\frac{1}{Tf^2}c_rL_{rs}c_s.
 \label{eq:reaction-drag-decomposition}
\end{align}
At fixed $T$ and $L_{rs}$, the first term gives a damping force $-\gamma_{\mathrm{reac}}u^\mu\nabla_\mu\phi$. The contraction $c_rL_{rs}c_s$ depends only on the symmetric part of $L$, denoted by $L_{\mathrm{sym}}\equiv(L+L^{\mathsf T})/2$. The fluctuation--dissipation relation $N=2L_{\mathrm{sym}}$ and SK positivity then give $\gamma_{\mathrm{reac}}=c_rN_{rs}c_s/(2Tf^2)\geq0$ for $T>0$. The second term captures how the fluid's chemical potentials feed back on the inflaton. In the perturbation analysis, the current derivatives defined above also account for the variation of $L_{rs}$ with temperature and affinity.

\paragraph{Stochastic description.}

A Hubbard--Stratonovich transformation of the imaginary quadratic terms in the parent action in Eq.~\eqref{eq:parent-action} introduces Gaussian reaction noise $w_r$ and transverse diffusion noise $I_a^\mu$. At fixed average fields, both noises have zero mean and their mixed covariance vanishes in the present truncation. Their local covariances are 
\begin{align}
 \langle w_r(x) w_s(x')\rangle
 &=N_{rs}\frac{\delta^{(4)}(x-x')}{\sqrt{-g(x')}},
 \nonumber\\
 \langle I_a^\mu(x)I_b^\nu(x')\rangle
 &=2T\sigma_{(ab)}\Delta^{\mu\nu}
 \frac{\delta^{(4)}(x-x')}{\sqrt{-g(x')}}.
 \label{eq:event-noise}
\end{align}
The coefficients and projector are evaluated at $x$. The stochastic contributions to the scalar, fluid-energy and charge equations are
\begin{align}\label{eq:parent-stochastic-sources}
 \xi_\phi&=-\frac{c_r}{f}w_r,\qquad
 \xi_\rho=\frac{c_r}{f}u^\mu\nabla_\mu\phi\, w_r,\qquad
 \xi_{n_a}=-\nu_{ra}w_r-\nabla_\mu I_a^\mu.
\end{align}
The shared reaction noise $w_r$ correlates the three sources through $N_{rs}$, and $I_a^\mu$ contributes through the diffusive charge current.

The equation of state $P(T,\mu_a)$, the reaction transport matrix $L_{rs}(T,\{\mathcal A_s/T\})$ and the conductivity matrix $\sigma_{ab}$ are constitutive inputs to be determined by microscopic matching for quantitative applications.

\section{Inflationary background and perturbations with a thermal fluid}
\label{sec:cosmology}

As a concrete application of the SK EFT, we analyze the warm-inflation model introduced in Sec.~\ref{sec:reactive-model}, in which the reactions responsible for inflaton dissipation also change the charge densities of the thermal fluid. Using the parent action in Eq.~\eqref{eq:parent-action}, we study how the resulting energy transfer and charge evolution affect the inflationary background and its perturbations. In Sec.~\ref{sec:background}, we derive the homogeneous evolution of the inflaton, temperature and chemical potentials. We solve the gravitational constraints and analyze the coupled physical perturbations in Sec.~\ref{sec:second-variation}, then construct their quadratic SK action and derive the associated noise correlations in Sec.~\ref{sec:physical-fields}.

\subsection{Warm inflationary background}
\label{sec:background}

We first determine the homogeneous evolution of the inflaton, temperature and chemical potentials.
We consider a spatially flat Friedmann--Lema\^{\i}tre--Robertson--Walker (FLRW) background,
\begin{align}
 \dd s^2&=-\dd t^2+a^2(t)\dd\bm x^2,\qquad H\equiv\frac{\dot a}{a}.
 \label{eq:frw-metric}
\end{align}
Here $a(t)$ and $H$ are the background scale factor and Hubble rate, written without bars.
Bars denote background values of the other fields and constitutive functions, and dots denote derivatives with respect to cosmic time $t$.
We set the external sources to zero as in Sec.~\ref{sec:kms}, so Eq.~\eqref{eq:zero-source-response} gives $\dot{\bar\psi}_a=\bar\mu_a$.
The response fields have zero background values, but we retain their variations to derive the background equations.
For the homogeneous background, the reaction current and affinity are
\begin{align}
 \bar J_r&\equiv J_r(\bar T,\{\bar{\mathcal A}_s\}),\qquad
 \bar{\mathcal A}_r=\frac{c_r}{f}\dot{\bar\phi}+\nu_{ra}\bar\mu_a.
\end{align}
Specializing Eqs.~\eqref{eq:einstein-balance}--\eqref{eq:bath-balance} derived above to the spatially flat FLRW background gives
\begin{align}
 3M_{\mathrm{Pl}}^2H^2
 &=\frac12\dot{\bar\phi}^{\,2}+V(\bar\phi)+\bar\rho,\qquad
 -2M_{\mathrm{Pl}}^2\dot H=\dot{\bar\phi}^{\,2}+\bar h,
 \label{eq:background-einstein}\\
 \ddot{\bar\phi}+3H\dot{\bar\phi}
 &=-V'(\bar\phi)-\frac{c_r}{f}\bar J_r,\qquad
 \dot{\bar\rho}+3H\bar h
 =\dot{\bar\phi}\frac{c_r}{f}\bar J_r,\qquad
 \dot{\bar n}_a+3H\bar n_a=-\nu_{ra}\bar J_r.
 \label{eq:background-fluid-charge}
\end{align}
The fluid-energy equation follows from the Bianchi identity and the scalar equation. The same background current governs scalar energy loss, fluid heating and charge transfer. A reaction channel with $c_r=0$ can change the charge composition without directly sourcing the inflaton equation. The fluid energy balance is written for the total energy density $\bar\rho$, which includes chemical energy.

To obtain the evolution of $\bar T$ and $\bar\mu_a$ from these balances, we use the general equation of state $P(T,\mu_a)$, whose background value is $\bar P\equiv P(\bar T,\{\bar\mu_a\})$.
The charge susceptibility matrix describes how the charge densities respond to chemical potentials at fixed temperature and is defined by
\begin{align}
 \chi_{ab}&\equiv\frac{\partial^2P}{\partial\mu_a\partial\mu_b}.
 \label{eq:susceptibility}
\end{align}
We consider a thermodynamically stable warm background with $\bar s,\bar h>0$, including mixed temperature--charge derivatives in the stability conditions.
We assume that $T$ and $\mu_a$ can be expressed as smooth functions of $\rho$ and $n_a$ near the background solution.
The energy and charge balances then give
\begin{align}
 \bar\rho_T\dot{\bar T}+\bar\rho_{\mu_b}\dot{\bar\mu}_b &=-3H\bar h+\dot{\bar\phi}\frac{c_r}{f}\bar J_r,
 \label{eq:temperature-balance}\\
 \bar n_{a,T}\dot{\bar T}+\bar\chi_{ab}\dot{\bar\mu}_b &=-3H\bar n_a-\nu_{ra}\bar J_r,
 \label{eq:chemical-balance}
\end{align}
where subscripts denote partial derivatives at fixed values of the other thermal variables.
The bars on these coefficients indicate background evaluation after differentiation, for example $\bar\rho_T\equiv\left.\rho_T\right|_{T=\bar T,\,\mu_a=\bar\mu_a}$.
Inverting the density Jacobian formed by the coefficients on the left gives $\dot{\bar T}$ and $\dot{\bar\mu}_a$.
If the density Jacobian becomes singular, the thermal evolution calls for a different choice of independent variables or a separate analysis.

\paragraph{Warm slowroll regime.}
We now examine how energy transfer from the inflaton sustains the fluid during slowroll. For $\dot{\bar\phi}\ne0$, we characterize the background dissipation by
\begin{align}
 \bar\gamma_{\mathrm{eff}} &\equiv\frac{1}{\dot{\bar\phi}}\frac{c_r}{f}\bar J_r, \qquad
 Q\equiv\frac{\bar\gamma_{\mathrm{eff}}}{3H}.
 \label{eq:background-friction}
\end{align}
Here $Q$ compares the effective background friction, including chemical feedback, with the Hubble friction $3H$.
The effective friction $\bar\gamma_{\mathrm{eff}}$ therefore differs from the background value $\bar\gamma_{\mathrm{reac}}$ of the coefficient in Eq.~\eqref{eq:reaction-drag-decomposition}, which isolates the scalar-velocity term at fixed transport coefficients.
To identify a slowly evolving inflationary background, we use slowroll parameters
\begin{align}
 \epsilon_1&\equiv-\frac{\dot H}{H^2}
 =\frac{\dot{\bar\phi}^{\,2}+\bar h}{2M_{\mathrm{Pl}}^2H^2},\qquad \epsilon_2\equiv\frac{\dot\epsilon_1}{H\epsilon_1},
 \qquad
 \epsilon_T\equiv\frac{\dot{\bar T}}{H\bar T},
 \qquad
 \epsilon_h\equiv\frac{\dot{\bar h}}{H\bar h},
 \qquad
 \epsilon_Q\equiv\frac{\dot Q}{HQ}.
 \label{eq:epsilon-enthalpy}
\end{align}
Quasi-de Sitter expansion corresponds to $\dot{\bar\phi}^{\,2}+\bar h\ll M_{\mathrm{Pl}}^2H^2$.
We consider a warm background in which energy transfer from the inflaton approximately balances dilution of the fluid by expansion~\cite{Moss:2008yb}.
Neglecting the slow change of its energy density in Eq.~\eqref{eq:background-fluid-charge} gives
\begin{align}
 |\dot{\bar\rho}|&\ll H\bar h,
 \qquad \bar h\simeq Q\dot{\bar\phi}^{\,2}.
 \label{eq:warm-attractor-balance}
\end{align}
Combining this with Eq.~\eqref{eq:epsilon-enthalpy}, we obtain
\begin{align}
 \dot{\bar\phi}^{\,2}&\simeq\frac{2\epsilon_1M_{\mathrm{Pl}}^2H^2}{1+Q}, \qquad
 \bar h\simeq
 \frac{2\epsilon_1M_{\mathrm{Pl}}^2H^2Q}{1+Q}.
 \label{eq:warm-enthalpy}
\end{align}
If scalar acceleration is negligible, the inflaton equation reduces to
\begin{align}
 3H(1+Q)\dot{\bar\phi}&\simeq-V'(\bar\phi).
 \label{eq:slow-roll-scalar-balance}
\end{align}
These balances describe the slowly evolving background.
The leading perturbation equations below additionally assume slow variation of the transport coefficients and susceptibilities.
The dynamical stability of this background requires an analysis of the coupled inflaton, thermal and charge perturbations introduced in Sec.~\ref{sec:second-variation}.

\paragraph{Background screening.}
\label{sec:two-charge}

The reactions driven by the inflaton can build up charge densities whose chemical potentials oppose the scalar contribution to the affinities.
This feedback can reduce the effective reaction force on the inflaton, an effect we call background screening~\cite{Drewes:2023khq,Berghaus:2025dqi}.
To determine this background screening, we first consider a quasistationary chemical state in which charge production approximately balances dilution by expansion.
Eq.~\eqref{eq:chemical-balance} then becomes
\begin{align}
 3H\bar n_a+\nu_{ra}\bar J_r&\simeq0.
 \label{eq:background-screening-balance}
\end{align}
At fixed $\bar T,H,\dot{\bar\phi}$, these equations determine $\bar\mu_a$ and hence the mean reaction force $ c_r\bar J_r/f$.
For illustration, consider one charge ($a=1$) and one reaction ($r=1$) in the linear regime, with $\bar n_1=\bar\chi_{11}\bar\mu_1$ and $\bar J_1=(\bar L_{11}/\bar T)\bar{\mathcal A}_1$.
Treating $\bar\chi_{11}$ and $\bar L_{11}$ as constants, Eq.~\eqref{eq:background-screening-balance} gives
\begin{align}
 \nu_{11}\bar\mu_1&\simeq
 -\frac{\nu_{11}^2\bar L_{11}/\bar T}{3H\bar\chi_{11}+\nu_{11}^2\bar L_{11}/\bar T}
 \frac{c_1}{f}\dot{\bar\phi}.
\end{align}
The minus sign shows that the chemical term $\nu_{11}\bar\mu_1$ opposes the inflaton contribution in $\bar{\mathcal A}_1=(c_1/f)\dot{\bar\phi}+\nu_{11}\bar\mu_1$.
This partial cancellation reduces the magnitude of the reaction current and the associated inflaton friction relative to the case $\bar\mu_1=0$.\footnote{For $H,\bar T,\bar\chi_{11},\bar L_{11}>0$ and $\nu_{11}\ne0$, the coefficient multiplying $-(c_1/f)\dot{\bar\phi}$ lies between zero and one, so the cancellation is partial.
The friction is reduced from its value at $\bar\mu_1=0$ by the factor $3H\bar\chi_{11}/(3H\bar\chi_{11}+\nu_{11}^2\bar L_{11}/\bar T)$, which also lies between zero and one.
These bounds follow from the assumptions of this linear example.}

\subsection{Scalar perturbations and gravitational constraints}
\label{sec:second-variation}

We next study small scalar perturbations of the inflaton, fluid and charges about the homogeneous solution.
Expanding the action in Eq.~\eqref{eq:parent-action} gives their linear coupled equations and Gaussian noise.
We then solve the gravitational constraints to express the nondynamical metric variables in terms of the remaining perturbations.
Coupled inflaton--radiation perturbations with thermal fluctuations have been studied in Refs.~\cite{Hall:2003zp,Montefalcone:2023pvh,Laine:2024wyv,Laine:2025rll}.
We use $\delta$ for perturbations of the average fields about the background, while $\Delta$ continues to denote response fields.
To organize the expansion, we denote the left-hand sides of the deterministic field equations by
\begin{align}
 \mathcal E_g^{\mu\nu}&\equiv T_\phi^{\mu\nu}+T_{\mathrm h}^{\mu\nu} -M_{\mathrm{Pl}}^2G_{\mathrm E}^{\mu\nu},\qquad \mathcal
 E_\phi\equiv\Box\phi-V'(\phi)-\frac{c_r}{f}J_r,\qquad
 \mathcal E_a\equiv\nabla_\mu j_a^\mu+\nu_{ra}J_r.
 \label{eq:charge-residual}
\end{align}
Using the response fields in Eqs.~\eqref{eq:zero-source-response} and \eqref{eq:zero-source-affinity}, we integrate the charge-current term by parts and write the deterministic action as
\begin{align}
 S_{\mathrm{det}}&=\int\dd^4x\sqrt{-g}\left[
 \frac12\mathcal E_g^{\mu\nu}G_{\Delta\mu\nu}
 +\mathcal E_\phi\Phi_\Delta
 -\mathcal E_a\left(\psi_{a\Delta}
 +X_\Delta^\mu\partial_\mu\psi_a\right)\right].
 \label{eq:deterministic-residual-action}
\end{align}
Using $X_\Delta^\mu\partial_\mu\bar\psi_a=\bar\mu_aX_\Delta^0$ and taking the spatial Fourier transform, we obtain the second order action
\begin{align}
 S_{\mathrm{det}}^{(2)}&=\int\dd t\,a^3 \int\frac{\dd^3\bm k}{(2\pi)^3}\left[ \frac12\delta\mathcal E_{g,-\bm
 k}^{\mu\nu}G_{\Delta\mu\nu,\bm k} +\delta\mathcal E_{\phi,-\bm k}\Phi_{\Delta,\bm k}
 -\delta\mathcal E_{a,-\bm k}
 (\psi_{a\Delta,\bm k}+\bar\mu_a X^0_{\Delta,\bm k})\right].
 \label{eq:quadratic-variation}
\end{align}
Here $\bm k$ is the comoving wave vector, $k\equiv|\bm k|$, and $\dd^3\bm k/(2\pi)^3$ is the Fourier integration measure.
The imaginary terms already contain two response fields, so their coefficients are evaluated on the background at this order.
At this order, the noise terms in Eq.~\eqref{eq:parent-action} depend on the response fields only through the linearized combinations $\mathcal A_{r\Delta}$ and $B_{a\Delta i}$. We retain $X_\Delta^\mu$ while solving the gravitational constraints.

We need to solve the gravitational constraints for the lapse and scalar shift.
In the standard scalar metric decomposition, spatially flat gauge on both branches gives
\begin{align}
 \dd s^2&=-(1+2\alpha)\dd t^2+2a\partial_iB\,\dd t\,\dd x^i +a^2\delta_{ij}\dd x^i\dd x^j,\qquad
 u_i\equiv a\partial_i v,
 \label{eq:flat-gauge}
\end{align}
where $\alpha$ is the lapse perturbation, $B$ the scalar shift potential and $v$ the fluid velocity potential.
{We define perturbations of the scalar component $q_{\mathrm{tot}}$ of the total momentum density  and the total energy-density $\rho_{\mathrm{tot}}$ through $\delta T_\phi{}^0{}_i+\delta T_{\mathrm h}{}^0{}_i=\partial_i\delta q_{\mathrm{tot}}$ and $\delta T_\phi{}^0{}_0+\delta T_{\mathrm h}{}^0{}_0=-\delta\rho_{\mathrm{tot}}$.
Expanding the scalar and ideal-fluid stress tensors in Eqs.~\eqref{eq:scalar-stress-definition} and \eqref{eq:fluid-stress} in the spatially flat gauge of Eq.~\eqref{eq:flat-gauge}, we obtain}
\begin{align}
 \delta q_{\mathrm{tot},\bm k}
 &=-\dot{\bar\phi}\delta\phi_{\bm k}
 +a(\bar\rho+\bar P)v_{\bm k},\qquad
 \delta\rho_{\mathrm{tot},\bm k}
 =\dot{\bar\phi}\dot{\delta\phi}_{\bm k}
 -\dot{\bar\phi}^{\,2}\alpha_{\bm k}
 +V'(\bar\phi)\delta\phi_{\bm k}
 +\delta\rho_{\bm k}.
 \label{eq:constraint-matter}
\end{align}
For $k\ne0$, the momentum and Hamiltonian constraints determine the lapse and shift:
\begin{align}
 \alpha_{\bm k}&=-\frac{\delta q_{\mathrm{tot},\bm k}} {2M_{\mathrm{Pl}}^2H},\qquad \frac{k^2}{a}B_{\bm k}=
 \frac{\delta\rho_{\mathrm{tot},\bm k}}{2M_{\mathrm{Pl}}^2H} +3H\alpha_{\bm k}.
 \label{eq:constraint-solutions}
\end{align}
These expressions include the slowroll gravitational terms.
In the action, each response field is paired with an equation for the average fields.
We substitute the lapse and shift solutions into the deterministic part of the quadratic action. The corresponding response constraints involve the adjoint operators obtained by integration by parts. The noise covariance of the physical variables is obtained in Sec.~\ref{sec:physical-fields} by projecting the stochastic sources of Sec.~\ref{sec:reactive-model} onto the reduced equations.

\paragraph{Thermal and reaction perturbations.}
The remaining equations couple temperature and charge fluctuations to the inflaton through the equation of state and the reaction currents.
Linearizing the pressure $P(T,\mu_a)$ and the thermodynamic relations in Eq.~\eqref{eq:thermodynamics} about the background gives
\begin{align}
 \delta\rho_{\bm k}&=\bar\rho_T\delta T_{\bm k} +\bar\rho_{\mu_a}\delta\mu_{a,\bm k},\qquad \delta n_{a,\bm
 k}=\bar n_{a,T}\delta T_{\bm k} +\bar\chi_{ab}\delta\mu_{b,\bm k},\qquad
 \label{eq:thermodynamic-perturbations}
 \delta P_{\bm k}=\bar s\delta T_{\bm k}+\bar n_a\delta\mu_{a,\bm k}.
\end{align}
Using the current derivatives introduced in Eq.~\eqref{eq:differential-current}, the reaction-rate perturbations are
\begin{align}
 \delta J_{r,\bm k}&=\bar J_{r,T}\delta T_{\bm k} +\bar J_{r,\mathcal A_s}\delta\mathcal A_{s,\bm k},\qquad
 \delta\mathcal A_{r,\bm k}= \frac{c_r}{f}(\dot{\delta\phi}_{\bm k}-\alpha_{\bm k}\dot{\bar\phi})
 +\nu_{ra}\delta\mu_{a,\bm k},
 \label{eq:reaction-perturbation}
\end{align}
where $\bar J_{r,T}$ and $\bar J_{r,\mathcal A_s}$ are evaluated at $\bar T,\{\bar{\mathcal A}_s\}$ after differentiation.
The temperature and affinity derivatives hold the dimensional affinities and $T$ fixed, respectively, and include the dependence of the transport coefficients.
Thus temperature and chemical fluctuations change the reaction currents, which feed back into the scalar and fluid equations.

\subsubsection{Physical perturbation variables}
After eliminating the nondynamical metric variables, we express the remaining perturbations in variables that separate total curvature from the relative motion and composition of the fluid.
On the regular thermodynamic branch described above, with $\dot{\bar\phi}\ne0$, we define the total comoving curvature and scalar--fluid relative velocity by
\begin{align}
 \mathcal R_{\bm k}&\equiv-H\frac{\delta q_{\mathrm{tot},\bm k}} {\dot{\bar\phi}^{\,2}+\bar\rho+\bar P},\qquad
 \mathcal S_{\bm k}\equiv-3H\left(\frac{\delta\phi_{\bm k}}{\dot{\bar\phi}} +av_{\bm k}\right).
 \label{eq:physical-curvature-sound}
\end{align}
$\mathcal S_{\bm k}$ is a nonadiabatic degree of freedom supporting fluid sound at finite momentum.
We describe fluctuations in the charge sector using the ratios $\mu_a/T$ and their background derivatives, which remain regular at zero chemical potential.
The relative perturbations of these ratios are defined by
\begin{align}
 C_{a,\bm k}&\equiv\delta\qty(\frac{\mu_a}{T})_{\bm k} -\frac{\partial_t(\bar\mu_a/\bar T)}{\dot{\bar\phi}}\delta\phi_{\bm k}.
 \label{eq:relative-chemical-fields}
\end{align}
The subtraction removes a common background time shift.
Inverting the curvature and velocity definitions, we obtain
\begin{align}
 \delta\phi_{\bm k}=\frac{\dot{\bar\phi}}{H}\mathcal R_{\bm k} -\frac{\dot{\bar\phi}(\bar\rho+\bar P)}
 {3H(\dot{\bar\phi}^{\,2}+\bar\rho+\bar P)}\mathcal S_{\bm k},\qquad
 av_{\bm k}=-\frac{\mathcal R_{\bm k}}H -\frac{\dot{\bar\phi}^{\,2}} {3H(\dot{\bar\phi}^{\,2}+\bar\rho+\bar
 P)}\mathcal S_{\bm k}. \label{eq:scalar-in-physical-fields}
\end{align}
On singular branches, different variables or a reduced set of independent variables should be used.

To keep a direct relation to the original inflaton equation in Eq.~\eqref{eq:charge-residual} and its response field, we use the curvature variable built from the inflaton perturbation when writing the action.
This inflaton-clock curvature is
\begin{align}
 \mathcal R_{\phi,\bm k}&\equiv\frac{H}{\dot{\bar\phi}}\delta\phi_{\bm k}
 =\mathcal R_{\bm k}
 -\frac{\bar h}{3(\dot{\bar\phi}^{\,2}+\bar h)}\mathcal S_{\bm k}.
 \label{eq:inflaton-clock-curvature}
\end{align}
Equation~\eqref{eq:inflaton-clock-curvature} recovers the total curvature from this variable and the relative velocity.
The two curvatures coincide once $\mathcal S$ has decayed, but their common late-time value can depend on the earlier coupled evolution of $\mathcal R_\phi$ and $\mathcal S$~\cite{Bastero-Gil:2014jsa,Bastero-Gil:2019rsp}.
We denote the corresponding response fields by $\mathcal R_{\phi\Delta}$, $\mathcal S_\Delta$ and $C_{a\Delta}$ and fix their normalization through the transformed action. 
These response fields satisfy
\begin{align}
 \mathcal R_{\phi\Delta,\bm k}-3\mathcal S_{\Delta,\bm k}
 &=-\frac{\dot{\bar\phi}}H\phi_{\Delta,\bm k}
 =-\frac{\dot{\bar\phi}^{\,2}}{H^2}
 \big[(\mathcal R_\phi)_{+,\bm k}-(\mathcal R_\phi)_{-,\bm k}\big].
 \label{eq:clock-response-normalization}
\end{align}
The last equality is linearized on a common background, and its sign follows Eq.~\eqref{eq:quadratic-variation}.
The stochastic forces and their covariance are transformed with the equations, preserving the coupled dynamics.

\paragraph{Thermal and reaction-current reconstruction.}
To close the equations in these variables, we express $\delta T$ and $\delta\mu_a$ in terms of $\mathcal R_\phi$, $\mathcal S$ and $C_a$ before substituting them into the reaction currents.
The linearized fluid momentum equation obtained from Eq.~\eqref{eq:bath-balance} supplies the pressure perturbation, while Eqs.~\eqref{eq:relative-chemical-fields} and \eqref{eq:thermodynamic-perturbations} give the thermal conversion.
The resulting relations are
\begin{align}
 \frac{\delta T_{\bm k}}{\bar T}
 &=\frac{\dot{\mathcal R}_{\phi,\bm k}}H
 +\frac{\dot{\mathcal S}_{\bm k}}{3H}
 + {\epsilon_T}\mathcal R_{\phi,\bm k}
 +\left[1+ {\frac{\epsilon_h}{3}}
 +\frac{\epsilon_1\dot{\bar\phi}^{\,2}}
 {3(\dot{\bar\phi}^{\,2}+\bar h)}\right]\mathcal S_{\bm k}
 -\frac{\bar T}{\bar h}\bar n_b C_{b,\bm k},
 \label{eq:general-thermal-conversion}\\
 \delta\mu_{a,\bm k}
 &=\bar T C_{a,\bm k}+\bar\mu_a\frac{\delta T_{\bm k}}{\bar T}
 +\frac{\bar T}{H}\partial_t\qty(\frac{\bar\mu_a}{\bar T})\mathcal R_{\phi,\bm k}.
 \label{eq:general-chemical-conversion}
\end{align}
These relations include the background time derivatives and hold for a general equation of state.
Substituting these relations into Eq.~\eqref{eq:reaction-perturbation} gives the reaction-current perturbations in the chosen variables as
\begin{align}
 \delta J_{r,\bm k}
 &=\frac{\dot{\bar J}_r}{H}\mathcal R_{\phi,\bm k}
 +\frac{\bar T \bar J_{r,T}+\bar J_{r,\mathcal A_s}\bar{\mathcal A}_s}{H}
 \dot{\mathcal R}_{\phi,\bm k}
 +\frac{\bar T \bar J_{r,T}
 +\bar J_{r,\mathcal A_s}\nu_{sb}\bar\mu_b}{3H}
 \dot{\mathcal S}_{\bm k}
 \nonumber\\
 &\quad+\left\{\left[\bar T \bar J_{r,T}
 +\bar J_{r,\mathcal A_s}\nu_{sb}\bar\mu_b\right]
 \left[1+ {\frac{\epsilon_h}{3}}
 +\frac{\epsilon_1\dot{\bar\phi}^{\,2}}
 {3(\dot{\bar\phi}^{\,2}+\bar h)}\right]
 -\frac{\epsilon_1\dot{\bar\phi}\bar h}
 {3f(\dot{\bar\phi}^{\,2}+\bar h)}
 \bar J_{r,\mathcal A_s}c_s\right\}\mathcal S_{\bm k}
 \nonumber\\
 &\quad+\bar T\left\{\bar J_{r,\mathcal A_s}\nu_{sb}
 -\frac{\bar n_b}{\bar h}\left[\bar T \bar J_{r,T}
 +\bar J_{r,\mathcal A_s}\nu_{sd}\bar\mu_d\right]\right\}C_{b,\bm k}.
 \label{eq:general-physical-current}
\end{align}
The chain rule gives the background derivative appearing here as
\begin{align}
 \dot{\bar J}_r&=\bar J_{r,T}\dot{\bar T}
 +\bar J_{r,\mathcal A_s}\left[
 \frac{c_s}{f}\ddot{\bar\phi}+\nu_{sb}\dot{\bar\mu}_b\right].
 \label{eq:background-current-chain-rule}
\end{align}

For the thermodynamic coefficients, we regard $P$ and $\mu_a/T$ as functions of the densities $(\rho,n_a)$ and define
\begin{align}
 P_\rho\equiv\left.\frac{\partial P}{\partial\rho}\right|_n,\qquad
 P_a\equiv\left.\frac{\partial P}{\partial n_a}\right|_{\rho,\,n_{c\neq a}},\qquad
 q_{a\rho}\equiv\left.\frac{\partial(\mu_a/T)}{\partial\rho}\right|_n,\qquad
 q_{ab}\equiv\left.\frac{\partial(\mu_a/T)}{\partial n_b}\right|_{\rho,\,n_{c\neq b}}.
 \label{eq:physical-density-derivatives}
\end{align}
These derivatives follow by inverting the linear thermodynamic relations \eqref{eq:thermodynamic-perturbations}, holding all other density coordinates fixed.
Bars denote their evaluation at $\bar\rho,\bar n_a$.
For each reaction, the associated changes in fluid energy and charge densities enter the relative perturbations of $\mu_a/T$ through
\begin{align}
 \mathcal Q_{ar}
 &\equiv-\left[\left.\frac{\partial(\mathcal A_r/T)}{\partial n_a}\right|_{\rho,\,n_{b\ne a}}\right]_{\mathrm{bg}}
 =\frac{\dot{\bar\phi}c_r}{f}\bar q_{a\rho}-\nu_{rb}\bar q_{ab}.
 \label{eq:relative-reaction-projection}
\end{align}
In taking derivatives, we fix the energy and charge transfers per reaction, $(c_r/f)u^\mu\nabla_\mu\phi$ and $-\nu_{rb}$.

\paragraph{Leading-order perturbation equations.}
At leading order in slowroll, we neglect gravitational corrections in the equations of this paragraph.
The numerical spectra include first-order slowroll corrections.

The constraints are imposed before this reduction.
Using Eqs.~\eqref{eq:constraint-solutions}, \eqref{eq:physical-curvature-sound} and \eqref{eq:epsilon-enthalpy}, their metric contributions can be written as
\begin{align}
 \alpha_{\bm k}&=\epsilon_1\mathcal R_{\bm k},\qquad
 \frac{k^2}{a}B_{\bm k}
 =\frac{\delta\rho_{\mathrm{tot},\bm k}}{2M_{\mathrm{Pl}}^2H}
 +3\epsilon_1H\mathcal R_{\bm k}.
 \label{eq:constraint-slow-roll-counting}
\end{align}
The lapse is explicitly proportional to $\epsilon_1$.
To establish the suppression of the density contribution to the shift, we use the definition of $\mathcal R_\phi$ in Eq.~\eqref{eq:inflaton-clock-curvature} and the background scalar equation in Eq.~\eqref{eq:background-fluid-charge}.
Together with $\alpha_{\bm k}=\epsilon_1\mathcal R_{\bm k}$, these give the relation
\begin{align}
 \delta\rho_{\mathrm{tot},\bm k}
 &=\dot{\bar\phi}^{\,2}
 \left[
 \frac{\dot{\mathcal R}_{\phi,\bm k}}H
 -3(1+Q)\mathcal R_{\phi,\bm k}
 +\epsilon_1
 (\mathcal R_{\phi,\bm k}-\mathcal R_{\bm k})
 \right]
 +\delta\rho_{\bm k}.
\end{align}
On the quasistationary warm background, Eq.~\eqref{eq:warm-enthalpy} then gives, at leading order,
\begin{align}
 \frac{\delta\rho_{\mathrm{tot},\bm k}}
 {2M_{\mathrm{Pl}}^2H^2}
 &\simeq\epsilon_1
 \left[
 \frac{\dot{\mathcal R}_{\phi,\bm k}}{H(1+Q)}
 -3\mathcal R_{\phi,\bm k}
 +\frac{Q}{1+Q}
 \frac{\delta\rho_{\bm k}}{\bar h}
 \right].
\end{align}
Thus both contributions to $k^2B_{\bm k}/a$ are suppressed by $\epsilon_1$ when $\mathcal R_{\bm k}$, $\mathcal R_{\phi,\bm k}$, $\dot{\mathcal R}_{\phi,\bm k}/H$ and $\delta\rho_{\bm k}/\bar h$ remain finite in the slowroll expansion.
In particular, the factors $1/(1+Q)$ and $Q/(1+Q)$ do not introduce an enhancement at large $Q$.
Under this condition we can take the decoupling limit of gravitational contributions and explicit slow background variations in Eq.~\eqref{eq:general-physical-current}.
The physical momentum $k/a$ continues to redshift, and the reaction currents still respond to temperature and affinity perturbations.

Neglecting the background energy-density and charge-density derivatives in Eq.~\eqref{eq:background-fluid-charge} gives
\begin{align}
 3H\bar h&=\dot{\bar\phi}\frac{c_r}{f}\bar J_r,
 \qquad
 3H\bar n_a=-\nu_{ra}\bar J_r.
 \label{eq:leading-warm-slice-balances}
\end{align}
Using these balances, we express the linearized scalar, fluid and charge equations \eqref{eq:scalar-balance}--\eqref{eq:bath-balance} in terms of $\mathcal R_\phi$, $\mathcal S$ and $C_a$.
Here $c_{\mathrm{ad}}^2\equiv\bar P_\rho+\bar n_b\bar P_b/\bar h$ is the adiabatic sound speed squared at fixed $n_a/s$.
{The stochastic forces $\xi_{\mathcal R_\phi}$, $\xi_{\mathcal S}$ and $\xi_{C_a}$ are the forces in Eq.~\eqref{eq:parent-stochastic-sources} after the same field and equation transformations.
At leading order in the approximation specified above, the coupled evolution equations take the following form:}
\begin{align}
 \ddot{\mathcal R}_{\phi,\bm k}
 &=-3H\dot{\mathcal R}_{\phi,\bm k}
 -\frac{k^2}{a^2}\mathcal R_{\phi,\bm k}
 -\frac{H}{f\dot{\bar\phi}} c_r\delta J_{r,\bm k}
 +\xi_{\mathcal R_\phi,\bm k},
 \label{eq:general-clock-row}\\
 {\ddot{\mathcal S}_{\bm k}}
 &=-3H(2+\bar P_\rho)\dot{\mathcal S}_{\bm k}
 -\left[9H^2(1+\bar P_\rho)
 +c_{\mathrm{ad}}^2\frac{k^2}{a^2}\right]\mathcal S_{\bm k}+3(1-c_{\mathrm{ad}}^2)\frac{k^2}{a^2}\mathcal R_{\phi,\bm k}
 \notag\\
 &\quad+\frac{3H}{\bar h}
 \left[\frac{\dot{\bar\phi}c_r}{f}\bar P_\rho
 -\nu_{rb}\bar P_b+\frac{\bar h c_r}{f\dot{\bar\phi}}\right]
 \delta J_{r,\bm k}-\frac{3H\bar T}{\bar h}\frac{k^2}{a^2}
 \bar P_b\bar\sigma_{bd}C_{d,\bm k}
 +{\xi_{\mathcal S,\bm k}},
 \label{eq:general-sound-row}\\
 \dot C_{a,\bm k}
 &=-3HC_{a,\bm k}
 -\bar h\bar q_{a\rho}
 (\dot{\mathcal S}_{\bm k}+3H\mathcal S_{\bm k})
 +\mathcal Q_{ar}
 \delta J_{r,\bm k}
 \nonumber\\
 &\quad-\frac{k^2}{a^2H}
 \left[\bar h\bar q_{a\rho}
 +\bar n_b
 \bar q_{ab}\right]
 (\mathcal R_{\phi,\bm k}+\mathcal S_{\bm k}/3)
 -\bar T\frac{k^2}{a^2}
 \bar q_{ab}
 \bar\sigma_{bd}C_{d,\bm k}+\xi_{C_a,\bm k},
 \label{eq:general-charge-row}
\end{align}
At this order, the temperature, chemical-potential and reaction-current perturbations are reconstructed algebraically as
\begin{align}
 \frac{\delta T_{\bm k}}{\bar T}
 &\simeq\frac{\dot{\mathcal R}_{\phi,\bm k}}H
 +\frac{\dot{\mathcal S}_{\bm k}}{3H}+\mathcal S_{\bm k}
 -\frac{\bar T}{\bar h}\bar n_b C_{b,\bm k},
 \label{eq:leading-thermal-closure}\\
 \delta\mu_{a,\bm k}
 &\simeq\bar T C_{a,\bm k}
 +\bar\mu_a\frac{\delta T_{\bm k}}{\bar T},
 \label{eq:leading-chemical-closure}\\
 {\delta J_{r,\bm k}}
 &{\simeq\frac{\bar T\bar J_{r,T}
 +\bar J_{r,\mathcal A_s}\bar{\mathcal A}_s}{H}
 \dot{\mathcal R}_{\phi,\bm k}+\frac{\bar T\bar J_{r,T}
 +\bar J_{r,\mathcal A_s}\nu_{sd}\bar\mu_d}{3H}
 (\dot{\mathcal S}_{\bm k}+3H\mathcal S_{\bm k})}
 \nonumber\\
 &{\quad+\bar T\left\{\bar J_{r,\mathcal A_s}\nu_{sb}
 -\frac{\bar n_b}{\bar h}
 \left[\bar T\bar J_{r,T}
 +\bar J_{r,\mathcal A_s}\nu_{sd}\bar\mu_d\right]\right\}C_{b,\bm k}.}
 \label{eq:leading-current-closure}
\end{align}
Eqs.~\eqref{eq:leading-thermal-closure}--\eqref{eq:leading-current-closure} reconstruct the thermal perturbations and reaction currents from $\mathcal R_\phi$, $\mathcal S$, $C_a$ and their derivatives once the background, equation of state and transport functions are specified.
No independent equations for $\delta T$, $\delta\mu_a$ or $\delta J_r$ are required in the deterministic evolution.
The stochastic forces are correlated projections of the parent reaction and diffusion noise, with their covariances given in Sec.~\ref{sec:physical-fields}.

\subsection{Quadratic action and noise correlations}
\label{sec:physical-fields}

We now write the quadratic SK action for $\mathcal R_\phi$, $\mathcal S$ and $C_a$, retaining the gravitational contributions and background time dependence. Each response field multiplies the deterministic left-hand side of its corresponding equation. For physical perturbations $A$ and $B$, the differential operator $\mathcal D_{AB,k}$ gives the contribution of $B$ to the equation for $A$. The terms quadratic in the response fields contain $N_{AB,k}$, which specifies the covariance between the stochastic forces on the right-hand sides of the equations for $A$ and $B$. These covariances follow from projecting the reaction and diffusion noise of Sec.~\ref{sec:reactive-model} onto the reduced equations.
The action takes the form
\begin{align}
 S^{(2)}&=\int\dd t\,a^3\int\frac{\dd^3\bm k}{(2\pi)^3}\Bigg\{ \mathcal R_{\phi\Delta,-\bm k}\left[
 \mathcal D_{\mathcal R_\phi\mathcal R_\phi,k}\mathcal R_{\phi,\bm k} +\mathcal D_{\mathcal R_\phi\mathcal S,k}\mathcal S_{\bm k}
 +\mathcal D_{\mathcal R_\phi C_a,k}C_{a,\bm k}\right]
 \nonumber\\
 &\qquad\qquad\qquad\qquad +\mathcal S_{\Delta,-\bm k}\left[ \mathcal D_{\mathcal S\mathcal R_\phi,k}\mathcal R_{\phi,\bm k} +\mathcal D_{\mathcal
 S\mathcal S,k}\mathcal S_{\bm k} +\mathcal D_{\mathcal S C_a,k}C_{a,\bm k}\right]
 \nonumber\\
 &\qquad\qquad\qquad\qquad+ C_{a\Delta,-\bm k}\left[ \mathcal D_{C_a\mathcal R_\phi,k}\mathcal R_{\phi,\bm k} +\mathcal D_{C_a\mathcal
 S,k}\mathcal S_{\bm k} +{\mathcal D_{C_aC_b,k}C_{b,\bm k}}\right]
 \nonumber\\
 &\qquad+\frac{\mathrm i}{2}\bigg[ N_{\mathcal R_\phi\mathcal R_\phi,k}\mathcal R_{\phi\Delta,-\bm k}\mathcal R_{\phi\Delta,\bm k}
 +N_{\mathcal S\mathcal
 S,k}\mathcal S_{\Delta,-\bm k}\mathcal S_{\Delta,\bm k}+N_{C_aC_b,k}C_{a\Delta,-\bm k}C_{b\Delta,\bm k}
 \nonumber\\
 &\qquad+2N_{\mathcal R_\phi\mathcal S,k}\mathcal R_{\phi\Delta,-\bm k}\mathcal S_{\Delta,\bm k} +2N_{\mathcal R_\phi C_a,k}\mathcal R_{\phi\Delta,-\bm k}C_{a\Delta,\bm k} +2N_{\mathcal S C_a,k}\mathcal
 S_{\Delta,-\bm k}C_{a\Delta,\bm k} \bigg]\Bigg\}.
 \label{eq:physical-gaussian-structure}
\end{align}
With the current derivatives in Eq.~\eqref{eq:differential-current}, the balance equations \eqref{eq:scalar-balance}--\eqref{eq:bath-balance}, subject to the constraints \eqref{eq:constraint-solutions}, give $\mathcal D$ in the physical variables \eqref{eq:physical-curvature-sound}, \eqref{eq:relative-chemical-fields} and \eqref{eq:inflaton-clock-curvature}.
The thermal and reaction-current perturbations are specified by Eqs.~\eqref{eq:general-thermal-conversion}, \eqref{eq:general-chemical-conversion} and \eqref{eq:general-physical-current}.
The explicit evolution equations to first order in slowroll are collected in Appendix~\ref{app:full-physical-evolution}.

\paragraph{Gaussian noise.}
{The imaginary part of Eq.~\eqref{eq:physical-gaussian-structure} fixes the noise matrix of the physical variables.
With the response vector
\begin{align}
 \boldsymbol q_{\Delta,\bm k}
 &\equiv(\mathcal R_{\phi\Delta,\bm k},\mathcal S_{\Delta,\bm k},
 C_{1\Delta,\bm k},\ldots,C_{m\Delta,\bm k})^{\mathsf T},
 \label{eq:physical-noise-vectors}
\end{align}
we collect its coefficients in
\begin{align}
 \boldsymbol N_k
 &\equiv\begin{pmatrix}
 N_{\mathcal R_\phi\mathcal R_\phi,k}&N_{\mathcal R_\phi\mathcal S,k}&N_{\mathcal R_\phi C_b,k}\\
 N_{\mathcal S\mathcal R_\phi,k}&N_{\mathcal S\mathcal S,k}&N_{\mathcal S C_b,k}\\
 N_{C_a\mathcal R_\phi,k}&N_{C_a\mathcal S,k}&N_{C_aC_b,k}
 \end{pmatrix}.
 \label{eq:physical-force-noise-matrix}
\end{align}
Here $m$ is the number of charges.
This matrix is real symmetric and positive semidefinite, with its coefficients evaluated on the background.
The matrix $\boldsymbol N_k$ fixes the Gaussian law of $\boldsymbol\xi_{\bm k}\equiv (\xi_{\mathcal R_\phi,\bm k},\xi_{\mathcal S,\bm k}, \xi_{C_1,\bm k},\ldots,\xi_{C_m,\bm k})^{\mathsf T}$ through the Hubbard--Stratonovich identity\footnote{
In the ideal-stress theory, the noises satisfy $\dot{\bar\phi}\xi_\phi+\xi_\rho=0$ by Eq.~\eqref{eq:parent-stochastic-sources}, which makes $\boldsymbol N_k$ singular.
The Gaussian integral is restricted to the covariance support, and $\boldsymbol N_k^{-1}$ denotes the inverse on this subspace.
}
\begin{align}
 e^{-\operatorname{Im}S^{(2)}}
 =
 \frac{1}{\mathcal Z_\xi}\int\!\mathcal D\boldsymbol\xi\,
 \exp\Bigg\{-\int\dd t\,a^3(t)\frac{\dd^3\bm k}{(2\pi)^3}\Big[
 \frac12\boldsymbol\xi_{-\bm k}^{\mathsf T}
 \boldsymbol N_k^{-1}\boldsymbol\xi_{\bm k}
 +\mathrm i\boldsymbol q_{\Delta,-\bm k}^{\mathsf T}
 \boldsymbol\xi_{\bm k}\Big]\Bigg\}.
 \label{eq:physical-noise-hs}
\end{align}
The forces therefore satisfy
\begin{align}
 \left\langle\boldsymbol\xi_{\bm k}(t)\right\rangle_\xi=0,\qquad
 \left\langle\boldsymbol\xi_{\bm k}(t)
 \boldsymbol\xi_{\bm k'}^{\mathsf T}(t')\right\rangle_\xi
 =(2\pi)^3\delta^{(3)}(\bm k+\bm k')
 \frac{\delta(t-t')}{a^3(t')}
 \boldsymbol N_k(t).
 \label{eq:physical-force-covariance}
\end{align}
Every entry of $\boldsymbol N_k$ is thus the corresponding force covariance.
After Eq.~\eqref{eq:physical-noise-hs}, integration over the response fields imposes the deterministic rows of Eq.~\eqref{eq:physical-gaussian-structure} with right-hand sides $\xi_{\mathcal R_\phi}$, $\xi_{\mathcal S}$ and $\xi_{C_a}$, respectively.
Projecting the sources in Eq.~\eqref{eq:parent-stochastic-sources}, using the density derivatives and reaction projection in Eqs.~\eqref{eq:physical-density-derivatives} and \eqref{eq:relative-reaction-projection}, gives the forces in Eqs.~\eqref{eq:general-clock-row}--\eqref{eq:general-charge-row}:
\begin{align}
 \xi_{\mathcal R_\phi,\bm k}
 &=-\frac{Hc_r}{f\dot{\bar\phi}}w_{r,\bm k},\label{eq:R-physical-forces}
 \\
 \xi_{\mathcal S,\bm k}
 &=\frac{3H}{\bar h}
 \left[\frac{\dot{\bar\phi}c_r}{f}\bar P_\rho-\nu_{rb}\bar P_b
 +\frac{\bar h c_r}{f\dot{\bar\phi}}\right]
 w_{r,\bm k}
 -\frac{3H}{\bar h}\bar P_b[\nabla_\mu I_b^\mu]_{\bm k},
 \label{eq:general-physical-forces}\\
 \xi_{C_a,\bm k}
 &=\bar q_{a\rho}\xi_{\rho,\bm k}+\bar q_{ab}\xi_{n_b,\bm k}=\mathcal Q_{ar}w_{r,\bm k}
 -\bar q_{ab}[\nabla_\mu I_b^\mu]_{\bm k}.
 \label{eq:physical-charge-force}
\end{align}
The $\delta\phi$ subtraction in $C_a$ contributes velocity terms to its first-order equation, but no direct white-noise force, since $\dot{\delta\phi}$ is kinematic.}

\section{Primordial power spectrum and environmental response}
\label{sec:spectrum}

Sound response~\cite{Graham:2009bf,Laine:2025rll} and charge dynamics~\cite{Dejrah:2026ltr} have been studied in warm inflation; here their evolution, energy transfer and noise follow from the constrained SK action of Sec.~\ref{sec:cosmology}.
Together with an initial covariance, the coupled operators and noise correlations summarized in Sec.~\ref{sec:physical-fields} determine the final correlators.
Sec.~\ref{sec:power} uses this system to compute spectra and compare the effects of suppressing chemical and sound perturbations.
Sec.~\ref{sec:reduction} examines the response hierarchy required for a local approximation sometimes assumed in literature.

\subsection{Primordial power spectrum}
\label{sec:power}

We define primordial spectra in the superhorizon adiabatic limit where nonadiabatic transients and noise-driven time dependence have decayed:\footnote
{
The fluid and inflaton clocks align through Eq.~\eqref{eq:inflaton-clock-curvature} as the fluid contributions decay.
For example, when local thermal white noise gives the slowest-decaying contribution, a residual clock mismatch is proportional to $a^{-3/2}$.
See Ref.~\cite{Moss:2008yb} for an early discussion and Refs.~\cite{Mirbabayi:2022cbt,Laine:2024wyv} for explicit analyses.
}
\begin{align}
 \langle\mathcal R_{\bm k}\mathcal R_{\bm k'}\rangle
 &=(2\pi)^3\delta^{(3)}(\bm k+\bm k')P_{\mathcal R}(k),\qquad
 \mathcal P_{\mathcal R}(k)=\frac{k^3}{2\pi^2}P_{\mathcal R}(k).
 \label{eq:power-definition}
\end{align}
The average includes both the initial ensemble and the subsequent Hubbard--Stratonovich noise.
At a pivot $k_*$, the scalar tilt and tensor-to-scalar ratio are defined by
\begin{align}
 n_s-1&\equiv\left.\frac{\partial\ln\mathcal P_{\mathcal R}(k)}
 {\partial\ln k}\right|_{k=k_*},\qquad
 r\equiv\frac{\mathcal P_t(k_*)}{\mathcal P_{\mathcal R}(k_*)}.
 \label{eq:tilt-tensor-definitions}
\end{align}
A star denotes background evaluation at horizon crossing of the pivot mode.
For the numerical scalar spectra, we solve the first-order slowroll equations in Appendix~\ref{app:full-physical-evolution}, using the evolving charged background and the gravitational constraints and correlated noise of Sec.~\ref{sec:physical-fields}.
For tensors we use the leading Bunch--Davies vacuum result,\footnote{
The general formulation in Sec.~\ref{sec:parent} includes shear of the fluid as well, which modifies the tensor modes \cite{Qiu:2021ytc,Klose:2022rxh,Biondini:2026wfy}.
}
\begin{align}
 \mathcal P_t(k_*)&\simeq\frac{2H_*^2}{\pi^2M_{\mathrm{Pl}}^2}.
 \label{eq:vacuum-tensor-power}
\end{align}

{For an estimate of the scale dependence, consider a smooth, stable quasistationary thermal branch.
The background rates defined in Eq.~\eqref{eq:epsilon-enthalpy}, together with the chemical variations $H^{-1}\partial_t(\bar\mu_a/\bar T)$, organize the slow evolution of the coupled system.
Differentiating Eq.~\eqref{eq:warm-enthalpy} and using the Bunch--Davies reference $\mathcal P_{\mathcal R}^{(0)}=H^4/(4\pi^2\dot{\bar\phi}^{\,2})$ gives the first-order decomposition}
\begin{align}
 n_s-1\simeq-2\epsilon_{1*}-\epsilon_{2*}
 +\frac{Q_*\epsilon_{Q*}}{1+Q_*}
 +\left.\frac{\dd}{\dd\ln k}
 \ln\frac{\mathcal P_{\mathcal R}(k)}{\mathcal P_{\mathcal R}^{(0)}(k)}\right|_{k=k_*},\qquad
 r\simeq\frac{16\epsilon_{1*}}{1+Q_*}
 \frac{\mathcal P_{\mathcal R}^{(0)}(k_*)}{\mathcal P_{\mathcal R}(k_*)}.
 \label{eq:leading-coupled-ratio}
\end{align}
Temperature and chemical variations enter the scale dependence of $\mathcal{P}_\mathcal{R}/\mathcal{P}_\mathcal{R}^{(0)}$.
In the full time-dependent system they enter {the thermal conversion \eqref{eq:general-thermal-conversion}, the reaction-current variation \eqref{eq:general-physical-current}, and the physical force covariance \eqref{eq:physical-force-covariance}.}
{For example, $H^{-1}\partial_t\ln(\bar T/H)\simeq\epsilon_T+\epsilon_1$ to first order in slowroll.}
The constitutive laws and background balances \eqref{eq:temperature-balance}--\eqref{eq:chemical-balance} fix the remaining transport and chemical variations; these are not all independent slowroll inputs.
The background dissipation ratio $Q$ includes mean chemical screening through Eq.~\eqref{eq:background-friction}, but does not by itself encode the response of the relative charge perturbations or sound.

\paragraph{Numerical illustration.}
We take $V=\lambda\phi^4$ and a QCD sphaleron-heated bath, motivated by Refs.~\cite{Berghaus:2025dqi,ORamos:2025uqs,Broadberry:2025ggb,Dejrah:2026ltr}.
The $\mathcal N_*$ counts e-folds backwards from the end of inflation $\epsilon_1\sim 1$, while the power spectrum is read out before $\epsilon_1\sim 1$ but after it freezes in superhorizon.
For each $f$ and $\mathcal N_*$, we fix $\lambda$ by matching $\mathcal P_{\mathcal R}(k_*)\simeq2.105\times10^{-9}$ at $k_*=0.05\,\mathrm{Mpc}^{-1}$~\cite{Planck:2018jri}; the background evolution determines the Hubble parameter and slowroll parameters.
We choose the Bunch--Davies vacuum for the inflaton as well as set the fluid and charge fluctuations to zero in spatially flat gauge, which follows literature~\cite{Rodrigues:2025neh,Dejrah:2026ltr}.
Thermal fluctuations are generated by noises during the subsequent evolution.

As in Ref.~\cite{Dejrah:2026ltr}, we evolve five axial quark densities $n_q$ ($q=u,d,s,c,b$) and one hard gauge-helicity moment $n_h$ with conjugate bias $\mu_h$.
Treating this kinetic moment as a locally advected thermodynamic variable is a constitutive assumption of this illustration; we omit higher helicity moments.
For the bath we adopt
\begin{align}
 P(T,\{\mu_a\})
 &=\frac{\pi^2g_*}{90}T^4
 +\frac12\sum_{a=q,h}\chi_{aa}(T)\mu_a^2,
 \qquad \rho=3P,
 \label{eq:illustrative-pressure}
\end{align}
where $N_c=3$, $g_*=106.75$, $\chi_{qq}=N_cT^2/3$, $\chi_{hh}=(N_c^2-1)T^2/3$, and the cross susceptibilities vanish.
Here $g_*$ is the relativistic ideal-gas SM value above the electroweak scale~\cite{Dejrah:2026ltr}, and we use the susceptibility normalization of the same reference.
We adopt this quadratic pressure at finite bias on the convex domain $\sum_a\chi_{aa}\mu_a^2<2\pi^2g_*T^4/45$.
The chemical contribution in Eq.~\eqref{eq:illustrative-pressure} enters both the background stress and its perturbations.
{For spatial diffusion, we neglect cross conductivities for these
axial charges, motivated by the approximately diagonal conserved
quark-flavor transport at high temperature and vanishing chemical
potentials~\cite{Arnold:2000dr,Fotakis:2024hmz,Chen:2013tra}.
For each charge, including helicity, we adopt the phenomenological conductivity matrix of Eq.~\eqref{eq:diffusion} as $\sigma_{aa}=0.1\chi_{aa}/T$.
Here we assume rapid current relaxation, $\tau_q\ll H^{-1}$.
In a kinetic description, the second-order current-relaxation time obeys $\tau_q\ge3\sigma_{qq}/\chi_{qq}$~\cite{Ghiglieri:2018dgf}.}

We adopt the phenomenological chirality-relaxation estimate used in
Refs.~\cite{Berghaus:2025dqi,Dejrah:2026ltr}, $\Gamma_{\mathrm{ch},q}=0.01\,y_q^2(T)T$, where $y_q(T)$ is the running Yukawa coupling of quark flavor $q$.
We evaluate the couplings $\alpha_{\mathrm g}(T)$ and $y_q(T)$ entering the reaction rates using one-loop SM running in $\overline{\mathrm{MS}}$, neglecting quark mixing and setting the renormalization scale equal to the bath temperature.
The boundary inputs are the central values of the gauge couplings and running fermion masses at the Z-boson mass given in
Ref.~\cite{Xing:2007fb}.
We use the label $23$ for the effective $2\leftrightarrow3$ channel in which scattering processes such as $gg\leftrightarrow ggg$ relax the net helicity of hard gluons.
The reaction noise strengths are $N_{\mathrm{sph}}=\kappa(\alpha_{\mathrm g})(N_c\alpha_{\mathrm g})^5T^4,\ N_{23}=2\alpha_{\mathrm g}^3T^4,\ N_{\mathrm{ch},q}=2T\chi_{qq}\Gamma_{\mathrm{ch},q}$.
The choice $N_{23}=2\alpha_{\mathrm g}^3T^4$ is the fiducial hard-helicity mobility of Ref.~\cite{Dejrah:2026ltr}, based on the structure discussed in Ref.~\cite{Broadberry:2025ggb}. 
For the sphaleron normalization, we use the next-to-leading-logarithmic prescription of Ref.~\cite{Dejrah:2026ltr}, $\kappa(\alpha_{\mathrm g})=0.28\,W_0\!\left[{e^{3.041}}\sqrt{{8\pi}/{9\alpha_{\mathrm g}}}\right]$, where $W_0$ is the principal branch of the Lambert $W$ function.
See also Ref.~\cite{Moore:2010jd} for the computation of the sphaleron rate.
Both the sphaleron normalization and the thermal running include all six quark flavors, while we evolve only the five axial quark densities specified above.

The sphaleron, hard-helicity and chirality-relaxation event currents are\footnote{{
We use the equilibrium value of $N_{\mathrm{sph}}$ in $J_{\mathrm{sph}}$ even when the background affinity satisfies $|\bar{\mathcal A}_{\mathrm{sph}}|/\bar T\gg1$.
Ref.~\cite{Ghoshal:2026qio} recently discussed this extrapolation in warm inflation with QCD plasma.
}}
\begin{align}
 J_{\mathrm{sph}}
 {=}\frac{N_{\mathrm{sph}}}{2T}
 \left(
 \frac{u^\mu\nabla_\mu\phi}{f}
 {+}2\sum_q\mu_q{-}\frac{\mu_h}{\alpha_{\mathrm g}(T)}
 \right),\quad
 J_{23}
 {=}\frac{N_{23}}{2T}
 \left(
 \alpha_{\mathrm g}(T)\frac{u^\mu\nabla_\mu\phi}{f}{-}\mu_h
 \right),\quad
 J_{\mathrm{ch},q}
 {=}\chi_{qq}\Gamma_{\mathrm{ch},q}\mu_q.
 \label{eq:illustrative-event-currents}
\end{align}
{Eq.~\eqref{eq:reaction-affinity-first-law} pairs the forces $u^\mu\nabla_\mu\phi/f$, $\mu_q$ and $\mu_h$ with the currents $J_\phi=J_{\mathrm{sph}}+\alpha_{\mathrm g}(T)J_{23}$, $J_q=2J_{\mathrm{sph}}+J_{\mathrm{ch},q}$ and $J_h=-J_{\mathrm{sph}}/\alpha_{\mathrm g}(T)-J_{23}$, respectively.}
{To evaluate $J_{r,T}$ in Eqs.~\eqref{eq:reaction-perturbation} and \eqref{eq:conditional-current-composition}, we hold these forces fixed and differentiate the projected currents, including the running of $\alpha_{\mathrm g}(T)$ in both the event affinities and the projection coefficients.}
The same projections fix the correlated reaction noise through Eq.~\eqref{eq:event-noise} and \eqref{eq:parent-stochastic-sources}, including the sphaleron contribution $+w_{\mathrm{sph}}/\alpha_{\mathrm g}(T)$ to the stochastic source of the helicity balance, with strengths given in the previous paragraph. 
For comparison with the first arXiv version of Ref.~\cite{Dejrah:2026ltr}, we set its sphaleron-to-helicity matching coefficient $\zeta_h$ to unity and identify its $\alpha_s$ with our $\alpha_{\mathrm g}$.
Its Eqs.~(4.3) and (B.5) correspond in our notation to $J_h=-2J_{\mathrm{sph}}/\alpha_{\mathrm g}(T)-J_{23}$ and the helicity-density noise contribution $+2w_{\mathrm{sph}}/\alpha_{\mathrm g}(T)$, respectively.
This would also require replacing $-\mu_h/\alpha_{\mathrm g}(T)$ by $-2\mu_h/\alpha_{\mathrm g}(T)$ in $J_{\text{sph}}$ in Eq.~\eqref{eq:illustrative-event-currents} to satisfy the first law, although Eq.~(3.1) of that reference uses $-\mu_h/\alpha_{\mathrm g}(T)$.

\begin{figure}[!t]
 \centering
 \includegraphics[width=0.9\linewidth]
 {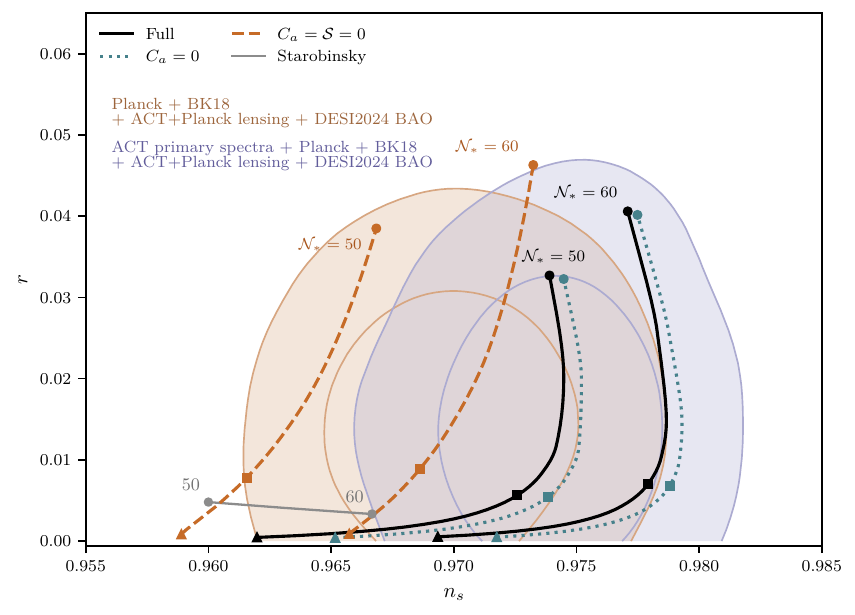}
 \caption{
 The $n_s$--$r$ curves, obtained by varying $f$ at $\mathcal N_*=50,60$, for Full and the two restricted systems defined in the text.
 The triangle, square, and circle markers correspond, respectively, to $f=(5,9,13)\times10^{12}\,\mathrm{GeV}$ for Full and $C_a=0$, and to $f=(3,6,9)\times10^{12}\,\mathrm{GeV}$ for $C_a=\mathcal S=0$.
 Orange shading gives the 68\% and 95\% reference
 $\Lambda$CDM+$r$ posteriors for Planck, BK18, ACT+Planck lensing
 and DESI2024 BAO
 \cite{AtacamaCosmologyTelescope:2025nti,BICEP:2021xfz};
 purple additionally includes ACT primary spectra.
 The gray Starobinsky segment \cite{Starobinsky:1980te} uses
 $n_s\simeq1-2/\mathcal{N}_*$ and $r\simeq12/\mathcal{N}_*^2$.
 }
 \label{fig:ver4-spectrum-overview}
\end{figure}

In Figure~\ref{fig:ver4-spectrum-overview}, the full system (``Full'') includes all six relative chemical perturbations and sound, and we compare Full with two artificial restrictions: $C_a=0$ and $C_a=\mathcal S=0$.
Each comparison restricts the bulk quadratic action to the remaining physical fields and their response partners, using the corresponding blocks of the initial and force covariances.
Figure~\ref{fig:ver4-spectrum-overview} shows that suppressing relative chemical perturbations shifts the curves modestly, whereas additionally suppressing sound produces larger changes.
{Note that, for the tensor-to-scalar ratio $r$ below the plotted range of Fig.~\ref{fig:ver4-spectrum-overview}, the curves of Full and $C_a=0$ turn back to the region favored by observation, which is consistent with Refs.~\cite{Berghaus:2025dqi,Dejrah:2026ltr}.}

The comparison shows that a small bath energy fraction does not by itself justify neglecting fluid perturbations in the curvature spectrum.
The quasistationary relations give
\begin{align}
 \frac{\bar\rho}{3M_{\mathrm{Pl}}^2H^2}
 &\simeq\frac{\epsilon_1Q}{2(1+Q)},
 \qquad
 \frac{\bar h}{3(\dot{\bar\phi}^{\,2}+\bar h)}
 \simeq\frac{Q}{3(1+Q)}.
 \label{eq:bath-energy-curvature-weights}
\end{align}
This shows that the coefficient of the relative-fluid term in Eq.~\eqref{eq:inflaton-clock-curvature} need not share the slowroll suppression of the bath energy fraction.

Figure~\ref{fig:ver4-spectrum-overview} also shows that increasing $f$ raises the tensor-to-scalar ratio $r$.
For comparison, cold inflation with the same quartic potential gives
\begin{align}
 n_s^{\mathrm{cold}}&\simeq1-\frac{3}{\mathcal{N}_*+1},
 \qquad
 r^{\mathrm{cold}}\simeq\frac{16}{\mathcal{N}_*+1}
\end{align}
at leading slowroll order.
The warm solutions have a smaller tensor-to-scalar ratio than this cold reference, reflecting the dissipative background and the bath contribution to scalar fluctuations in Eq.~\eqref{eq:leading-coupled-ratio}.
The increase of $r$ with $f$ is consistent with progressive decoupling from the bath, since the inflaton--bath coupling in Eq.~\eqref{eq:illustrative-event-currents} weakens as $f$ grows.
The additional response hierarchy required for a local adiabatic description is discussed in Sec.~\ref{sec:reduction}.

\subsection{Locality of the adiabatic effective description}
\label{sec:reduction}
\label{sec:environment-hierarchy}

We next examine when the environmental perturbations can be eliminated while keeping a local description of $\mathcal R_\phi$.
Trace-out generally produces a nonlocal influence functional, so a local approximation applied in Refs.~\cite{LopezNacir:2011kk,Creminelli:2023aly,Salcedo:2024smn} requires a separation between the environmental response scales and the spacetime scales relevant to the system, see also Ref.~\cite{Gangopadhyay:2026vdu} for memory effects in warm inflation, as well as Ref.~\cite{Burgess:2014eoa} for related scale-separation arguments.
The conditions stated in Sec.~\ref{sec:environment-fluid} specify the regime where a local fluid description with sound and charge perturbations applies.
Eliminating these perturbations requires a further hierarchy.
For the analytic hierarchy estimates below, we use the leading slowroll and quasistationary equations of Sec.~\ref{sec:second-variation} for the reactive fluid with ideal stress specified in Eq.~\eqref{eq:reactive-ideal-specialization}.
As an illustration, we consider smooth sources varying on a Hubble time or more slowly.
Faster variations require comparison with their own time scale.

\paragraph{Charge response.}
Substituting Eqs.~\eqref{eq:leading-thermal-closure}--\eqref{eq:leading-current-closure} into Eq.~\eqref{eq:general-charge-row}, we collect the terms proportional to the chemical perturbations into a local contribution and a diffusion contribution.
With $\mathcal R_\phi$ and $\mathcal S$ treated as prescribed sources, we consider the homogeneous deterministic charge equations:
\begin{align}
 \dot C_{a,\bm{k}}+\Gamma_{ab}C_{b,\bm{k}}
 +\frac{k^2}{a^2}D_{ab}C_{b,\bm{k}}&=0.
 \label{eq:charge-local-response}
\end{align}
The wavenumber-independent coefficient and diffusion coefficient are
\begin{align}
 \Gamma_{ab}
 &\equiv3H\delta_{ab}
 -\bar T\mathcal Q_{ar}
 \left\{
 \bar J_{r,\mathcal A_s}\nu_{sb}
 -\frac{\bar n_b}{\bar h}
 \left[
 \bar T\bar J_{r,T}
 +\bar J_{r,\mathcal A_s}\nu_{sd}\bar\mu_d
 \right]
 \right\},\qquad D_{ab}\equiv\bar T\bar q_{ac}\bar\sigma_{cb}.
 \label{eq:general-charge-generator}
\end{align}
The two terms in $\Gamma_{ab}$ describe dilution by expansion and the reaction-current response after thermal conversion.
The relaxation generated by the homogeneous operator in Eq.~\eqref{eq:charge-local-response} sets the time scales to be compared with variations of the prescribed sources.

Suppose that $\Gamma_{aa}>0$ and the charge mixing is small, $\sum_{b\ne a}|\Gamma_{ab}|\ll\Gamma_{aa}$ for every $a$, which is the case of the example above.
The response time of component $a$ is then approximately $\Gamma_{aa}^{-1}$.
Under the time-scale assumption stated above, sufficient conditions for a local derivative expansion of its driven response are
\begin{align}
 \frac{H}{\Gamma_{aa}},
 \frac{k^2}{a^2\Gamma_{aa}}\sum_b|D_{ab}|&\ll1,
 \label{eq:charge-local-time-hierarchy}
\end{align}
with no sum over $a$.
The first condition compares charge relaxation with the variation of the prescribed sources and background coefficients.
The second controls the spatial derivative expansion.
For an independently diffusing component, it reduces to $D_{aa}k^2/(a^2\Gamma_{aa})\ll1$: the diffusion distance during relaxation must be short compared with the physical wavelength.
If the nonzero ratios $D_{ab}/\Gamma_{aa}$ vary little over a Hubble time, redshifting reduces these spatial ratios approximately as $a^{-2}$ over the interval where the fluid description applies.
Note that the relaxation is assessed from the coupled charge equations when mixing $\Gamma_{ab}|_{a\neq b}$ is nonzero.

The term $3H\delta_{ab}$ in Eq.~\eqref{eq:general-charge-generator} describes dilution by expansion.
In the above numerical illustration, for example, the reaction contribution is negligible and $\Gamma_{aa}\simeq3H$.
Dilution alone therefore does not provide the temporal hierarchy in Eq.~\eqref{eq:charge-local-time-hierarchy}.
Local thermal equilibrium and stationary background balance alone do not ensure rapid charge relaxation.
These fast-response conditions are sufficient; cancellations between energy and charge transfer can also make the chemical contribution to the scalar--sound equations local.\footnote
{
Energy and charge transfer balance at the perturbative level if
$\nu_{ra}+(\bar n_a/\bar h)c_r\dot{\bar\phi}/f=0$,
as seen from the balance equations
\eqref{eq:scalar-charge-balances}--\eqref{eq:bath-balance}
and stochastic forces \eqref{eq:parent-stochastic-sources}.
For an ideal, quasistationary thermal fluid with negligible diffusion,
Eqs.~\eqref{eq:general-sound-row}--\eqref{eq:general-charge-row}
then allow $C_a$ to be removed algebraically from the bulk equations
in favor of $\mathcal R_\phi$, $\mathcal S$.
The balances \eqref{eq:leading-warm-slice-balances} enforce this alignment
whenever the same reaction dominates heating and charge production,
including the warm inflation with sphaleron but without additional
chirality violation \cite{Drewes:2023khq,Berghaus:2025dqi}.
}

\paragraph{Sound response.}
{Closing the currents in Eq.~\eqref{eq:general-sound-row} gives the leading-order sound operator, with $\mathcal R_\phi$ and $C_a$ treated as prescribed sources for simplicity.
Its homogeneous equation is}
\begin{align}
 \ddot{\mathcal S}_{\bm k}
 +\gamma_{\mathcal S}\dot{\mathcal S}_{\bm k}
 +\left(
 m_{\mathcal S}^2+c_{\mathrm{ad}}^2\frac{k^2}{a^2}
 \right)\mathcal S_{\bm k}&=0.
 \label{eq:sound-local-response}
\end{align}
The damping and zero-momentum coefficients are
\begin{align}
 {\gamma_{\mathcal S}}
 &{\equiv3H(2+\bar P_\rho)
 -\frac1{\bar h}
 \left[\frac{\dot{\bar\phi}c_r}{f}\bar P_\rho
 -\nu_{rb}\bar P_b+\frac{\bar h c_r}{f\dot{\bar\phi}}\right]
 \left(\bar T\bar J_{r,T}
 +\bar J_{r,\mathcal A_s}\nu_{sd}\bar\mu_d\right),}
 \label{eq:general-sound-damping}\\
 m_{\mathcal S}^2&\equiv3H\gamma_{\mathcal S}-9H^2.
 \label{eq:general-sound-scales}
\end{align}
The coefficient $\gamma_{\mathcal S}$ controls damping, but the decay time also depends on the restoring term.
Here $m_{\mathcal S}^2$ describes relative fluid motion of sound modes on the expanding, driven background; sound in static equilibrium is gapless \cite{Dubovsky:2011sj}.

For a stable diagonal response with $\gamma_{\mathcal S}>0$ and $m_{\mathcal S}^2>0$, a local derivative expansion of the driven solution for Hubble-time variations requires
\begin{align}
 \frac{H^2}{m_{\mathcal S}^2},
 \frac{H\gamma_{\mathcal S}}{m_{\mathcal S}^2},
 \frac{c_{\mathrm{ad}}^2k^2}{a^2m_{\mathcal S}^2}&\ll1.
 \label{eq:sound-mechanical-expansion}
\end{align}
The first two ratios compare the time-derivative terms with the restoring term; the third controls the spatial derivative expansion.
As an illustration at $k=0$, Eq.~\eqref{eq:sound-local-response} factorizes as $(\partial_t+\gamma_{\mathcal S}-3H)(\partial_t+3H)\mathcal S=0$, leading to two branches of the decay rates $3H$ and $\gamma_{\mathcal S}-3H$.
The stability $m_\mathcal{S}^2>0$ requires $\gamma_{\mathcal S}>3H$,\footnote
{
Although it indicates that the decay rate is never faster than $3H$, this alone does not mean nonlocality of curvature kernels. For frozen coefficients at $k=0$, the sound enters the clock and charge equations only through $\dot{\mathcal S}+3H\mathcal S$ as seen in Eqs.~\eqref{eq:general-clock-row}--\eqref{eq:leading-current-closure}, so the $3H$ branch drives neither $\mathcal R_\phi$ nor $C_a$. The $\gamma_\mathcal{S}-3H$ branch as well as prescribed sources control the locality in the kernel.
}
which implies
\begin{align}
 \frac{H\gamma_{\mathcal S}}{m_{\mathcal S}^2}
 &=\frac{\gamma_{\mathcal S}}
        {3(\gamma_{\mathcal S}-3H)}
 >\frac13 .
 \label{eq:sound-temporal-bound}
\end{align}
Therefore, the temporal derivative expansion of this diagonal sound response is not parametrically controlled for Hubble-time sources.
A fluid response lasting for a Hubble time also occurs in dissipative inflation coupled to a fluid \cite{Turiaci:2013dka}.

The spatial condition in Eq.~\eqref{eq:sound-mechanical-expansion} must be checked at the physical momenta relevant to scalar production; scalar freeze-out need not occur at $k/a=H$ \cite{LopezNacir:2011kk}.
The charge and sound response estimates above therefore show that local thermal equilibrium and strong background dissipation do not by themselves establish a local description of curvature alone.

\section{Conclusions}
\label{sec:conclusions}

We have developed a gravitational SK EFT for warm inflation with a reactive fluid carrying approximately conserved charges, building on gravitational SK and hydrodynamic constructions \cite{Haehl:2015foa,Haehl:2015uoc,Crossley:2015evo,Glorioso:2017fpd,Jensen:2017kzi,Landry:2020tbh,Hongo:2024brb,Lau:2024mqm}.
In the reactive specialization, the same currents determine scalar friction, fluid heating, charge transfer and correlated reaction noise.
Dynamical KMS symmetry gives the fluctuation--dissipation and Onsager--Casimir relations, while SK positivity constrains the allowed noise and dissipative coefficients.
Within the Gaussian-noise truncation, we derive the background and constrained linear perturbation equations from the common action, providing a basis for calculating spectra given the scalar-sector parameters, constitutive functions and initial state.

In an illustrative example with ideal fluid stress and sphaleron heating in Sec.~\ref{sec:power}, motivated by Refs.~\cite{Berghaus:2025dqi,Dejrah:2026ltr}, we evolve axial-charge and hard gauge-helicity perturbations together with the inflaton and sound.
With the scalar amplitude matched separately for each case, suppressing relative chemical perturbations shifts the tilt slightly, whereas additionally suppressing sound modes changes both the tilt and tensor-to-scalar ratio appreciably.

Section~\ref{sec:reduction} also examines the response hierarchy required for a local adiabatic description and derives the bound on the diagonal sound response in Eq.~\eqref{eq:sound-temporal-bound}.
On the regular branches considered there, the sound response can persist for a Hubble time \cite{Turiaci:2013dka}.
If this response contributes appreciably to curvature dynamics varying on the same time scale, the temporal separation needed for a controlled local expansion is absent.
Microscopic local equilibrium and strong background dissipation alone do not establish this hierarchy.

Extending the cosmological perturbation analysis to higher order would allow us to study how charge dynamics and correlated noise affect the bispectrum \cite{Turiaci:2013dka,Bastero-Gil:2014raa,Mirbabayi:2022cbt}.
Restoring viscous stress and its associated noise in the cosmological specialization would allow thermal tensor contributions to be computed within the same framework \cite{Qiu:2021ytc,Klose:2022rxh,Biondini:2026wfy}.
A further direction is to identify reactive-fluid setups that realize the response hierarchies underlying local single-clock descriptions of dissipative inflation \cite{LopezNacir:2011kk,Creminelli:2023aly,Salcedo:2024smn}, and to determine which sound and charge modes must remain dynamical outside that regime.

\section*{Acknowledgments}
We thank Perseas Christodoulidis, Masaru Hongo, Kohei Kamada, Maria Mylova, Ryo Namba, Atsuhisa Ota, Sota Sato, and Vaios Ziogas for their helpful discussions.
OpenAI ChatGPT and Anthropic Claude Code were used to assist with manuscript preparation, code development, and auxiliary analytic and numerical checks.
The authors take full responsibility for all arguments presented in this work.
T.N. is supported by JSPS KAKENHI Grant No. JP26K07101 and by the JGC-S Scholarship Foundation.
F.S. acknowledges support from JSPS KAKENHI Grant Number 26KJ0060.

\appendix

\section{First-order slowroll equations}
\label{app:full-physical-evolution}
{For the scalar spectra in Sec.~\ref{sec:power}, we evolve $\mathcal R_{\phi,\bm k}(t)$, $\mathcal S_{\bm k}(t)$ and $C_{a,\bm k}(t)$.
After imposing the constraints \eqref{eq:constraint-solutions}, we adopt the response-field normalization \eqref{eq:clock-response-normalization} and vary the quadratic action \eqref{eq:physical-gaussian-structure}.
The force covariance \eqref{eq:physical-force-covariance} and projections \eqref{eq:R-physical-forces}--\eqref{eq:physical-charge-force} fix the stochastic terms.}
We truncate the deterministic coefficients at first order of slowroll parameters, including gravitational contributions as well.
The constitutive functions and noise are evaluated on the full background solution of Eqs.~\eqref{eq:background-einstein}--\eqref{eq:background-fluid-charge} and \eqref{eq:temperature-balance}--\eqref{eq:chemical-balance}.

For a function $F$, we write
\begin{align}
 F_x^{(Y)}&\equiv\left.\frac{\partial F}{\partial x}\right|_Y,
 \label{eq:constitutive-derivative-notation}
\end{align}
where $Y$ specifies the quantities held fixed, while the density derivatives are defined in Eq.~\eqref{eq:physical-density-derivatives}.
Unvaried components of each coordinate set are held fixed, and bars denote background evaluation after differentiation.
Applying this rule to the currents gives
\begin{align}
 \bar J_{r,T}^{(\dot\phi,\mu/T)}
 \equiv\bar J_{r,T}
 +\frac{\bar J_{r,\mathcal A_s}\nu_{sb}\bar\mu_b}{\bar T},
 \quad
 \bar J_{r,T}^{(\mathcal A/T)}
 \equiv\bar J_{r,T}
 +\frac{\bar J_{r,\mathcal A_s}\bar{\mathcal A}_s}{\bar T},\quad
 \bar J_{r,\mu_b/T}^{(P,\dot\phi)}
 \equiv\bar T\bar J_{r,\mathcal A_s}\nu_{sb}
 -\frac{\bar T^2\bar n_b}{\bar h}
 \bar J_{r,T}^{(\dot\phi,\mu/T)}.
 \label{eq:conditional-current-composition}
\end{align}
Applying the same rule to the thermodynamic functions gives
\begin{align}
\bar\rho_T^{(\mu/T)}
 &\equiv\bar\rho_T+\frac{\bar\mu_b}{\bar T}\bar\rho_{\mu_b},
 \qquad
 \bar n_{a,T}^{(\mu/T)}
 \equiv\bar n_{a,T}+\frac{\bar\mu_b}{\bar T}\bar\chi_{ab}.
 \label{eq:app-thermal-responses}
 \\
 \bar\rho_{\mu_b/T}^{(P)}
 &\equiv\bar T\bar\rho_{\mu_b}
 -\frac{\bar T^2\bar n_b}{\bar h}\bar\rho_T^{(\mu/T)},
 \qquad
 \bar n_{a,\mu_b/T}^{(P)}
 \equiv\bar T\bar\chi_{ab}
 -\frac{\bar T^2\bar n_b}{\bar h}\bar n_{a,T}^{(\mu/T)}.
 \label{eq:app-isobaric-responses}
\end{align}
Time arguments are suppressed below.

The inflaton-clock curvature obeys
\begin{align}
&\ddot{\mathcal R}_{\phi,\bm k}
 +(\gamma_{\mathcal R_\phi}+\Delta\gamma_{\mathcal R_\phi})\dot{\mathcal R}_{\phi,\bm k}
 +\frac{k^2}{a^2}\mathcal R_{\phi,\bm k}
=\Bigg[
 -\frac{c_r}{3f\dot{\bar\phi}}\bar T\bar J_{r,T}^{(\dot\phi,\mu/T)}
 -\frac{\bar T\bar\rho_T^{(\mu/T)}-\bar h}{6M_{\mathrm{Pl}}^2H}
 \Bigg]\dot{\mathcal S}_{\bm k}
 \nonumber\\
 &\quad-\Bigg\{
 \frac{Hc_r}{f\dot{\bar\phi}}
 \Bigg[\left(1+\frac{\epsilon_h}{3}
 +\frac{\dot{\bar\phi}^{\,2}}{6M_{\mathrm{Pl}}^2H^2}\right)
 \bar T\bar J_{r,T}^{(\dot\phi,\mu/T)}
 -\frac{\dot{\bar\phi}\bar h}{6fM_{\mathrm{Pl}}^2H^2}
 \bar J_{r,\mathcal A_s}c_s\Bigg]
 +\frac{1}{2M_{\mathrm{Pl}}^2}
 \big(\bar T\bar\rho_T^{(\mu/T)}-\bar h\big)
 \Bigg\}\mathcal S_{\bm k}
 \nonumber\\
 &\quad-\Bigg[
 \frac{Hc_r}{f\dot{\bar\phi}}\bar J_{r,\mu_b/T}^{(P,\dot\phi)}
 +\frac1{2M_{\mathrm{Pl}}^2}\bar\rho_{\mu_b/T}^{(P)}
 \Bigg]C_{b,\bm k}
 +\xi_{\mathcal R_\phi,\bm k}.
 \label{eq:app-full-clock}
\end{align}
Define the scalar background rate by
\begin{align}
\epsilon_\phi&\equiv\frac{\ddot{\bar\phi}}{H\dot{\bar\phi}}=\frac{1}{2}\qty(1+\frac{\bar{h}}{\dot{\bar\phi}^{\,2}})(\epsilon_2-2\epsilon_1)-\frac{\bar{h}}{2\dot{\bar\phi}^{\,2}}\epsilon_h,
 \label{eq:app-scalar-slowroll-relation}
\end{align}
which is order of slowroll. The damping coefficients are
\begin{align}
&\gamma_{\mathcal R_\phi}\equiv3H+
 \frac{c_r\bar T}{f\dot{\bar\phi}}\bar J_{r,T}^{(\mathcal A/T)},\qquad
 \Delta\gamma_{\mathcal R_\phi}\equiv H(2\epsilon_\phi+\epsilon_1)
 +\frac{\bar T\bar\rho_T^{(\mu/T)}+\dot{\bar\phi}^{\,2}}{2M_{\mathrm{Pl}}^2H}.\label{eq:app-clock-coefficients}
\end{align}
Here $\gamma_{\mathcal R_\phi}$ contains Hubble damping and the reaction response, while $\Delta\gamma_{\mathcal R_\phi}$ accounts for background evolution and gravitational constraints.
The right-hand side couples curvature to sound and chemical fluctuations and to their common reaction noise.

Combining the fluid equations~\eqref{eq:scalar-charge-balances} and \eqref{eq:bath-balance} with Eq.~\eqref{eq:app-full-clock} to isolate $\ddot{\mathcal S}_{\bm k}$ gives
\begin{align}
&\ddot{\mathcal S}_{\bm k}
 +(\gamma_{\mathcal S}+\Delta\gamma_{\mathcal S})\dot{\mathcal S}_{\bm k}
 +\left[m_{\mathcal S}^{2}+\Delta m_{\mathcal S}^2
 +(c_{\mathrm{ad}}^2+\Delta c_{\mathcal S}^2)\frac{k^2}{a^2}\right]
 \mathcal S_{\bm k}=3(1-c_{\mathrm{ad}}^2)\frac{k^2}{a^2}\mathcal R_{\phi,\bm k}
 \nonumber\\
 &\qquad+3\Bigg[
 \frac1{\bar h}\left(\frac{\bar h c_r}{f\dot{\bar\phi}}+\frac{\dot{\bar\phi}c_r}{f}\bar P_\rho-\nu_{rd}\bar P_d\right)\bar T\bar J_{r,T}^{(\mathcal A/T)}
 +H(2\epsilon_\phi-\epsilon_h)
 -\frac{\bar P_b\dot{\bar n}_b}{\bar h}
 \nonumber\\
 &\qquad\qquad+\frac{\dot{\bar P}_\rho}{\bar h}\bar T\bar\rho_T^{(\mu/T)}
 +\frac{\dot{\bar P}_d}{\bar h}\bar T\bar n_{d,T}^{(\mu/T)}
 +\frac{1-c_{\mathrm{ad}}^2}{2M_{\mathrm{Pl}}^2H}
 \bigl[\bar T\bar\rho_T^{(\mu/T)}+\dot{\bar\phi}^{\,2}\bigr]
 \Bigg]\dot{\mathcal R}_{\phi,\bm k}
 \nonumber\\
 &\qquad+\Bigg[
 \frac{3H}{\bar h}\left(\frac{\bar h c_r}{f\dot{\bar\phi}}+\frac{\dot{\bar\phi}c_r}{f}\bar P_\rho-\nu_{rd}\bar P_d\right)
 \bar J_{r,\mu_b/T}^{(P,\dot\phi)}
 +\frac{3H\dot{\bar P}_\rho}{\bar h}\bar\rho_{\mu_b/T}^{(P)}
 +\frac{3H\dot{\bar P}_d}{\bar h}\bar n_{d,\mu_b/T}^{(P)}
 \nonumber\\
 &\qquad\qquad+\frac{3(1-c_{\mathrm{ad}}^2)}{2M_{\mathrm{Pl}}^2}\bar\rho_{\mu_b/T}^{(P)}
 -\frac{3H\bar T}{\bar h}\frac{k^2}{a^2}\bar P_d\bar\sigma_{db}
 \Bigg]C_{b,\bm k}
 +{\xi_{\mathcal S,\bm k}}.
 \label{eq:app-full-sound}
\end{align}
The damping, restoring and gradient coefficients are
\begin{align}
 &\Delta\gamma_{\mathcal S}\equiv2H(\epsilon_1+\epsilon_h)
 -\frac{\bar T}{\bar h}\big[\dot{\bar P}_\rho\bar\rho_T^{(\mu/T)}
 +\dot{\bar P}_d\bar n_{d,T}^{(\mu/T)}\big]
 -\frac{1-c_{\mathrm{ad}}^2}{2M_{\mathrm{Pl}}^2H}\bar T\bar\rho_T^{(\mu/T)},\\
 &\Delta m_{\mathcal S}^2\equiv3H\Delta\gamma_{\mathcal S}
 +3H^2\big[\bar P_\rho\epsilon_h+(\bar P_\rho-1)\epsilon_1\big]
 \nonumber\\
 &\quad-\frac{3H}{\bar h}
 \left(\frac{\bar h c_r}{f\dot{\bar\phi}}
 +\frac{\dot{\bar\phi}c_r}{f}\bar P_\rho-\nu_{rd}\bar P_d\right)
 \Bigg[\left(\frac{\epsilon_h}{3}
 +\frac{\dot{\bar\phi}^{\,2}}{6M_{\mathrm{Pl}}^2H^2}\right)
 \bar T\bar J_{r,T}^{(\dot\phi,\mu/T)}
 -\frac{\dot{\bar\phi}\bar h}{6fM_{\mathrm{Pl}}^2H^2}
 \bar J_{r,\mathcal A_s}c_s\Bigg],\\
 &\Delta c_{\mathcal S}^2\equiv-\frac{\bar T\bar P_b\bar\sigma_{bd}}{\bar h}
 \frac{\dd}{\dd t}\left(\frac{\bar\mu_d}{\bar T}\right).\label{eq:app-sound-coefficients}
\end{align}
The leading damping and restoring coefficients are those in Eqs.~\eqref{eq:general-sound-damping}--\eqref{eq:general-sound-scales}, with $c_{\mathrm{ad}}^2$ defined in Sec.~\ref{sec:second-variation}.
Their $\Delta$ corrections contain background drift and gravity.

\begin{samepage}
The relative chemical fluctuations satisfy the drift--diffusion equations
\begin{align}
&\dot C_{a,\bm k}
 +\left[\Gamma_{ab}+\Delta\Gamma_{ab}
 +\frac{k^2}{a^2}D_{ab}\right]C_{b,\bm k}
=
 -\frac{k^2}{a^2H}\bigl(\bar h\bar q_{a\rho}+\bar n_d\bar q_{ad}\bigr)\mathcal R_{\phi,\bm k}\notag\\
 &\qquad\quad+\Bigg[\frac{\mathcal Q_{ar}}H\bar T\bar J_{r,T}^{(\mathcal A/T)}
 +\frac{\dot{\bar q}_{a\rho}}H\bar T\bar\rho_T^{(\mu/T)}
 +\frac{\dot{\bar q}_{ad}}H\bar T\bar n_{d,T}^{(\mu/T)}
 -\frac{\bar q_{ab}\dot{\bar n}_b}H-\frac{\bigl(\bar h\bar q_{a\rho}+\bar n_d\bar q_{ad}\bigr)}{2M_{\mathrm{Pl}}^2H^2}
 \bigl[\bar T\bar\rho_T^{(\mu/T)}+\dot{\bar\phi}^{\,2}\bigr]
 \Bigg]\dot{\mathcal R}_{\phi,\bm k}
 \nonumber\\
 &\qquad\quad+\Bigg[-\bar h\bar q_{a\rho}
 +\frac{\bar T}{3H}\Bigg(
 \mathcal Q_{ar}\bar J_{r,T}^{(\dot\phi,\mu/T)}
 +\dot{\bar q}_{a\rho}\bar\rho_T^{(\mu/T)}
 +\dot{\bar q}_{ad}\bar n_{d,T}^{(\mu/T)}
 -\frac{\bar h\bar q_{a\rho}+\bar n_d\bar q_{ad}}{2M_{\mathrm{Pl}}^2H}
 \bar\rho_T^{(\mu/T)}\Bigg)
 \Bigg]\dot{\mathcal S}_{\bm k}
 \nonumber\\
 &\qquad\quad+\Bigg\{
 \mathcal Q_{ar}\Bigg[
 \left(1+\frac{\epsilon_h}{3}
 +\frac{\dot{\bar\phi}^{\,2}}{6M_{\mathrm{Pl}}^2H^2}\right)
 \bar T\bar J_{r,T}^{(\dot\phi,\mu/T)}
 -\frac{\dot{\bar\phi}\bar h}{6fM_{\mathrm{Pl}}^2H^2}
 \bar J_{r,\mathcal A_s}c_s\Bigg]
 \nonumber\\
 &\qquad\qquad+\bar T\big[
 \dot{\bar q}_{a\rho}\bar\rho_T^{(\mu/T)}
 +\dot{\bar q}_{ad}\bar n_{d,T}^{(\mu/T)}\big]
 -H\bar h\bar q_{a\rho}(3+\epsilon_1+\epsilon_h)
 -\frac{\bar T(\bar h\bar q_{a\rho}+\bar n_d\bar q_{ad})}
 {2M_{\mathrm{Pl}}^2H}\bar\rho_T^{(\mu/T)}
 \nonumber\\
 &\qquad\qquad+\frac{k^2}{3a^2H}
 \Big[(\bar q_{ab}\bar\sigma_{bd})
 \bar T\frac{\dd}{\dd t}\left(\frac{\bar\mu_d}{\bar T}\right)
 -(\bar h\bar q_{a\rho}+\bar n_d\bar q_{ad})\Big]
 \Bigg\}\mathcal S_{\bm k}
 +\xi_{C_a,\bm k}.
 \label{eq:app-full-charge}
\end{align}
Here $\Gamma_{ab}$ contains dilution and reaction response, and $D_{ab}$ controls diffusion, as defined in Eq.~\eqref{eq:general-charge-generator}.
The correction is defined by
\begin{align}
&\Delta\Gamma_{ab}\equiv-\dot{\bar q}_{a\rho}\bar\rho_{\mu_b/T}^{(P)}
 -\dot{\bar q}_{ad}\bar n_{d,\mu_b/T}^{(P)}
 +\frac{\bar h\bar q_{a\rho}+\bar n_d\bar q_{ad}}{2M_{\mathrm{Pl}}^2H}
 \bar\rho_{\mu_b/T}^{(P)}.
 \label{eq:app-response-increments}
\end{align}
Its first two terms arise from the evolving thermodynamic response, and the third from the gravitational constraint.
Curvature and sound drive the charge fluctuations through the terms on the right-hand side, together with the correlated force $\xi_{C_a}$.
\end{samepage}

The constitutive inputs and flat-gauge initial state of Sec.~\ref{sec:power}, mapped through Eqs.~\eqref{eq:physical-curvature-sound}--\eqref{eq:inflaton-clock-curvature} and \eqref{eq:general-thermal-conversion}, close the system.
The total comoving curvature follows from Eq.~\eqref{eq:inflaton-clock-curvature}.

\newpage
\addcontentsline{toc}{section}{References}
\bibliographystyle{setting/JHEP}
\bibliography{References}
\end{document}